\documentclass[aps, preprint, amsmath,amssymb,superscriptaddress]{revtex4-1}

\usepackage{graphicx}
\usepackage{xcolor,color}

\begin{document}

\title{Emergence of pseudogap from short-range spin-correlations in electron doped cuprates}

\author{F.\,Boschini$^\dagger$}
\email{boschini@phas.ubc.ca}
\address{Quantum Matter Institute, University of British Columbia, Vancouver, BC V6T 1Z4, Canada}
\address{Department of Physics $\&$ Astronomy, University of British Columbia, Vancouver, BC V6T 1Z1, Canada}
\author{M.\,Zonno}
\altaffiliation{These authors equally contributed}
\address{Quantum Matter Institute, University of British Columbia, Vancouver, BC V6T 1Z4, Canada}
\address{Department of Physics $\&$ Astronomy, University of British Columbia, Vancouver, BC V6T 1Z1, Canada}
\author{E.\,Razzoli}
\author{R.\,P.\,Day}
\address{Quantum Matter Institute, University of British Columbia, Vancouver, BC V6T 1Z4, Canada}
\address{Department of Physics $\&$ Astronomy, University of British Columbia, Vancouver, BC V6T 1Z1, Canada}
\author{M.\,Michiardi}
\address{Quantum Matter Institute, University of British Columbia, Vancouver, BC V6T 1Z4, Canada}
\address{Department of Physics $\&$ Astronomy, University of British Columbia, Vancouver, BC V6T 1Z1, Canada}
\affiliation{Max Planck Institute for Chemical Physics of Solids, N{\"o}thnitzer Stra{\ss}e 40, Dresden 01187, Germany}
\author{B.\,Zwartsenberg}
\author{P.\,Nigge}
\author{M.\,Schneider}
\address{Quantum Matter Institute, University of British Columbia, Vancouver, BC V6T 1Z4, Canada}
\address{Department of Physics $\&$ Astronomy, University of British Columbia, Vancouver, BC V6T 1Z1, Canada}
\author{E.\,H.\,da Silva Neto}
\address{Department of Physics, University of California, Davis, CA 95616, USA}
\author{A. Erb}
\address{Walther-Mei\ss ner-Institute for Low Temperature Research, Garching, 85748, Germany}
\author{S.\,Zhdanovich}
\author{A.\,K.\,Mills}
\author{G.\,Levy}
\address{Quantum Matter Institute, University of British Columbia, Vancouver, BC V6T 1Z4, Canada}
\address{Department of Physics $\&$ Astronomy, University of British Columbia, Vancouver, BC V6T 1Z1, Canada}
\author{C.\,Giannetti}
\address{Department of Mathematics and Physics, Universit\`{a} Cattolica del Sacro Cuore, Brescia, BS I-25121, Italy}
\address{Interdisciplinary Laboratories for Advanced Materials Physics (ILAMP),Universit\`{a} Cattolica del Sacro Cuore, Brescia I-25121, Italy}
\author{D.\,J.\,Jones}
\address{Quantum Matter Institute, University of British Columbia, Vancouver, BC V6T 1Z4, Canada}
\address{Department of Physics $\&$ Astronomy, University of British Columbia, Vancouver, BC V6T 1Z1, Canada}
\author{A.\,Damascelli}
\email{damascelli@physics.ubc.ca}
\address{Quantum Matter Institute, University of British Columbia, Vancouver, BC V6T 1Z4, Canada}
\address{Department of Physics $\&$ Astronomy, University of British Columbia, Vancouver, BC V6T 1Z1, Canada}

\date{\today}

\maketitle

\textbf{Electron interactions are pivotal for defining the electronic structure of quantum materials. In particular, the strong electron Coulomb repulsion is considered the keystone for describing the emergence of exotic and/or ordered phases of quantum matter as disparate as high-temperature superconductivity and charge- or magnetic-order. However, a comprehensive understanding of fundamental electronic properties of quantum materials is often complicated by the appearance of an enigmatic partial suppression of low-energy electronic states, known as the pseudogap. 
Here we take advantage of ultrafast angle-resolved photoemission spectroscopy to unveil the temperature evolution of the low-energy density of states in the electron-doped cuprate Nd$_{\text{2-x}}$Ce$_{\text{x}}$CuO$_{\text{4}}$, an emblematic system where the pseudogap intertwines with magnetic degrees of freedom.
By photoexciting the electronic system across the pseudogap onset temperature T*, we report the direct relation between the momentum-resolved pseudogap spectral features and the spin-correlation length with an unprecedented sensitivity. This transient approach, corroborated by mean field model calculations, allows us to establish the pseudogap in electron-doped cuprates as a precursor to the incipient antiferromagnetic order even when long-range antiferromagnetic correlations are not established, as in the case of optimal doping.}

\section*{Introduction}
In the presence of strong correlations, the interactions within and among various degrees of freedom often obfuscate the microscopic origin of exotic electronic phenomena \cite{FradkinReviewCupratesOrders,ReviewKeimerQuantumMat,ReviewBasov}. As a most prominent example, the interplay between intertwined orders \cite{PNAS_Davis_2013} continues to preclude a thorough understanding of the pseudogap (PG) phenomenon, a mysterious state of correlated matter by now notorious from systems as diverse as unconventional superconductors \cite{Timusk_RepProgPhys_1999,Hufner_RepProgPhys_2008,Vishik_RepProgPhys_2018}, dichalcogenides \cite{PseudogapCDWBorisenko,ReviewRossnagel}, and ultracold atoms \cite{PseudogapColdAtoms,RanderiaColdAtoms,Chen_ReviewPGcoldAtoms_2014}.
Broadly speaking, the PG in the condensed matter is associated with a partial suppression of the electronic spectral weight in the vicinity of the Fermi level ($\omega$=0), and evidence for the PG has been widely reported \cite{ReviewMott_MIT}. This behavior may be anticipated in the presence of long-range (or mesoscopic) order, \emph{e.g.} spin- or charge-order, which breaks the translational symmetry of the crystal: the loss of spectral weight in particular momentum-energy regions would be a simple consequence of the avoided crossings in the symmetry-reduced bandstructure \cite{ReviewKordyuk,NormanFermiArcModeling,PseudogapCDWBorisenko,ReviewRossnagel}. However, this argument may be unsatisfactory in the presence of strong electronic correlations and short-range orders with correlation length of few unit cells. 
Copper-oxide high-temperature superconductors are a paradigmatic example where the origin of the PG -- which presents different phenomenology for hole and electron doping -- is still debated, and a universal understanding has yet to emerge \cite{Timusk_RepProgPhys_1999,Hufner_RepProgPhys_2008,Vishik_RepProgPhys_2018,FradkinReviewCupratesOrders,ReviewKordyuk,NormanFermiArcModeling,KeimerReviewNature,ThermodynYBCONematic,ReviewElecDoped}.  

In the specific case of electron-doped cuprates, the PG is believed to bear a relation to the antiferromagnetic (AF) order which itself extends over a wide doping range \cite{ReviewElecDoped,NeutronNCCO_Nat2007,OpticalPG_NCCO,Matsui_ARPES_PRB2007,OnoseConductivity2004,Armitage_ARPES_PRL2001,Armitage_ARPES_PRL2002,Matsui_ARPES_PRL2005,PARK_ARPES_SF_PRB2013,Zimmers_Millis_EPL_PCCO}. As illustrated in Fig.\,\ref{Fig1}a, the PG is stable above the entire AF and superconducting (SC) domes, with its onset temperature indicated by T* as measured by spectroscopic and transport probes (orange shadow) \cite{ReviewElecDoped,OpticalPG_NCCO,Matsui_ARPES_PRB2007,OnoseConductivity2004,Zimmers_Millis_EPL_PCCO}. Scattering experiments on electron-doped cuprates have shown that the long-range AF order disappears when entering the narrow SC dome \cite{XrayNCCO_NatPhys2014,NeutronNCCO_Nat2007}, and that the commonly reported charge-order in cuprates does not exhibit a clear connection to the AF order \cite{EHdSN_ScienceCDW,EHdSN_ScienceAdvancesCDW}, although a coupling to dynamic magnetic correlations has been recently shown \cite{daSilvaNeto_NCCO_2018}. In addition, 3D collective charge modes, which may play a substantial role in mediating high-temperature superconductivity, have been reported \cite{Plasmon3D_LCCO}. 
In the presence of long-range AF order, \emph{i.e.} when the instantaneous spin-correlation length ($\xi_{spin}$) diverges at low temperature and a N\'{e}el temperature is defined, T* has been proposed to be a temperature crossover for which the quasiparticle de Broglie wavelength ($\lambda_\text{B} \approx v_\text{F}/\pi \text{T}$, where $v_\text{F}$ is the Fermi velocity) becomes comparable to $\xi_{spin}$ \cite{NeutronNCCO_Nat2007,TremblayTheorySF}. 
However, these considerations seem to fail where only short-range spin-fluctuations ($\xi_{spin}\lessapprox 50 a$, where $a$ is the unit cell size) are detected by inelastic neutron scattering \cite{NeutronNCCO_Nat2007}. 
In fact, for dopings where the long-range AF order disappears, \emph{i.e.} when the SC dome arises, the short-range $\xi_{spin}$ does not diverge at low temperature, preventing an unambiguous identification of a temperature crossover with $\lambda_\text{B}$. 
In addition, the underlying superconducting phase has been proposed to limit the development of $\xi_{spin}$ \cite{NeutronNCCO_Nat2007}. Finally, and most importantly, a momentum-resolved study connecting explicitly the PG spectral features and short-range AF correlations in electron-doped cuprates is still missing.
 
In an effort to tie together the observations of the PG and short-range AF correlations via a unique experimental approach, we performed a time- and angle-resolved photoemission (TR-ARPES) study of optimally doped Nd$_{\text{2-x}}$Ce$_{\text{x}}$CuO$_{\text{4}}$ (NCCO, T$_{\text{c}}\approx$24\,K, yellow arrow in Fig.\,\ref{Fig1}a), which is characterized by $\xi_{spin} \approx20 a$ for low temperatures (T$\approx$T$_{\text{c}}$) \cite{NeutronNCCO_Nat2007}.
TR-ARPES allows one to circumvent many of the challenges of a detailed temperature-dependent equilibrium ARPES study, such as surface degradation as well as coarse and uncorrelated sampling. As in standard pump-probe spectroscopy, a near-infrared pump pulse is used to perturb the system, with its relaxation studied by varying the temporal delay of a subsequent UV probe pulse. 
Due to the strong coupling of excited quasiparticles to underlying excitations, the energy released by the pump pulse is rapidly ($\approx$100\,fs) shared among all the degrees of freedom \cite{PerfettiPRL2007,ScienceDalConte,NatPhysDalConte,ReviewGiannetti,Hinton_TR_NCCO,Vishik_TR_LCCO}. 
After this initial relaxation, an effective electronic temperature T$_e$ may be defined at each point in time, allowing a temperature-dependent scan to be performed continuously and with remarkable detail \cite{NodePeakLanzara}. 
By applying this transient approach we reveal unambiguously the direct relation between momentum-resolved spectroscopic features of the PG and short range $\xi_{spin}$(T$_e$), as extracted from inelastic neutron scattering data reported in Ref.\,\cite{NeutronNCCO_Nat2007}. In addition, we identify T* as the crossover temperature above which the spin-fluctuation-induced spectral broadening exceeds the PG amplitude significantly, establishing the PG as a precursor of the underlying AF order. 

\section*{Results} 
\subsection*{Fermi surface mapping and modeling}
Figure\,\ref{Fig1}b displays the equilibrium Fermi surface mapping of NCCO acquired with 6.2\,eV probe pulsed-light. A tight-binding constant energy contour at $\omega$=0 \cite{TB_NCCO} (blue solid line), and the AF zone boundary (AFZB, red dashed line), are superimposed over the experimental Fermi surface. The intersection point between the tight-binding at $\omega$=0 and AFZB is commonly referred to as the hot-spot (HS), and coincides with the location where an AF-driven PG is expected to be particle-hole symmetric \cite{ReviewElecDoped}. 
In a mean field description, the commensurate $\textbf{q}$=($\pi$,$\pi$) folding of the Fermi surface is driven by a strong quasi-2D AF order in the copper-oxygen plane \cite{ReviewElecDoped}. The Green's function can then be written as \cite{NormanFermiArcModeling,Zimmers_Millis_EPL_PCCO,TremblayTheorySF}:
\begin{equation} \label{EQ:1}
  G^{-1}(\textbf{k},\omega)=\omega - \epsilon_{\textbf{k}} + i \eta - \frac{\Delta_{\text{PG}}^2}{\omega - \epsilon_{\textbf{k+q}} + i \Gamma},
\end{equation}
where $\epsilon_{\textbf{k}}$ is the bare energy dispersion, $\Delta_{\text{PG}}$ the AF-driven pseudogap spectroscopic amplitude, determined by the local Coulomb interaction and spin susceptibility \cite{TremblayTheorySF}, $\eta$ a generic quasiparticle broadening term, and $\Gamma$ the spin-fluctuation-induced spectral broadening term, which fills the pseudogap via the reduction of $\xi_{spin}$ \cite{TremblayTheorySF,PARK_ARPES_SF_PRB2013}. 
Using Eq.\,\ref{EQ:1} we can calculate the spectral function $A(\textbf{k},\omega)=-\frac{1}{\pi}\text{Im}[G(\textbf{k},\omega)]$ \cite{ReviewDamascelli} and compute the Fermi surface (Fig.\,\ref{Fig1}c), finding that it agrees well with our experimental data and previous ARPES mapping studies \cite{Armitage_ARPES_PRL2001,Armitage_ARPES_PRL2002}. We used $\Delta_{\text{PG}}$=$\eta$=$\Gamma$=85\,meV for simulation purposes, as evinced from our experimental data (see Fig.\,\ref{Fig1}d-f and Supplementary Information) and in agreement with previous optical and ARPES studies \cite{OpticalPG_NCCO,Matsui_ARPES_PRB2007,OnoseConductivity2004,Armitage_ARPES_PRL2001,Armitage_ARPES_PRL2002,Shen_NCCO_arXiv}.

\subsection*{Tracking the pseudogap spectral weight in an ultrafast fashion}
\begin{figure*}
\centering
\includegraphics[scale=0.81]{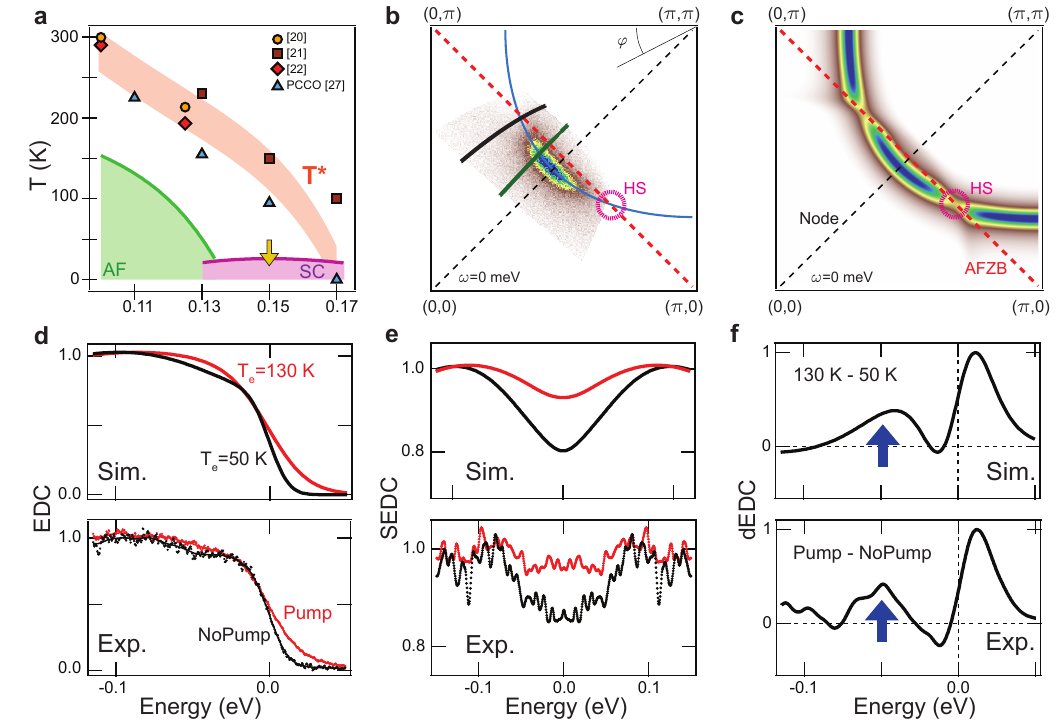}
\caption[Fig1]{\footnotesize Experimental strategy for tracking transient filling of the pseudogap.
\textbf{a}, Phase-diagram of the electron-doped cuprate Nd$_{\text{2-x}}$Ce$_{\text{x}}$CuO$_{\text{4}}$ showing the onset temperature of the pseudogap T* \cite{OpticalPG_NCCO,Matsui_ARPES_PRB2007,OnoseConductivity2004,Zimmers_Millis_EPL_PCCO}. The doping measured in this study is highlighted by the yellow arrow. 
\textbf{b}, Experimental Fermi surface of optimally-doped NCCO measured with 6.2-eV probe pulse, 10\,K base temperature. The integration window in energy is 20\,meV at the Fermi level. The solid blue line is a tight-binding constant energy contour at $\omega$=0 \cite{TB_NCCO}, the red dashed line the AF zone boundary (AFZB). The violet dotted circle encloses the hot-spot (HS). The black dashed line represents the nodal direction, and the green and the black solid lines the two momentum directions explored in this work, $\varphi \approx$39$^o$ and $\varphi \approx$26.5$^o$ respectively, where $\varphi$ is the angle between (0,$\pi$)-($\pi$,$\pi$) and the nodal direction. 
\textbf{c}, Simulated Fermi surface using Eq.\,\ref{EQ:1}, $\Delta_{\text{PG}}$=$\eta$=$\Gamma$=85\,meV (details in Supplementary Information). 
\textbf{d}, Energy distribution curves (EDCs) at the HS for 50\,K (black) and 130\,K (red). Top panel: EDCs simulated using Eq.\,\ref{EQ:1}, and $\Gamma$ equal to 85\,meV and 160\,meV for low and high temperature conditions, respectively. Bottom panel: experimental EDCs (dots, raw data; solid lines, smoothed data; HF regime).
\textbf{e}, Simulated (top) and experimental (bottom) symmetriezed EDCs (SEDCs). For T$_{\text{e}}$=130\,K the shortening of $\xi_{spin}$ leads to a filling-up of the PG (red curves).
\textbf{f}, Simulated (top) and experimental (bottom) differential EDCs (dEDCs), as defined in Eq.\,\ref{EQ:2}, where we demonstrate that a filling of the PG manifests as an increase of the photoemission intensity for $\omega \approx$-50\, meV (blue arrows).}
\label{Fig1} 
\end{figure*}

Before moving to a detailed analysis of our TR-ARPES data, we provide a more qualitative overview to illustrate the general concepts which underlie our experimental strategy.
Figure\,\ref{Fig1}d compares simulated and experimental energy distribution curves (EDCs) integrated along the momentum direction through HS for two (transient) electronic temperatures.
In an effort to reveal the fingerprint of the PG, whose evaluation may be challenging through the purely visual inspection of the data, we present the symmetrized EDCs (SEDCs, see Fig.\,\ref{Fig1}e). The symmetrization procedure removes any dependence of the photoemission signal on the Fermi-Dirac distribution function providing direct access to underlying modifications of the local density of states (DOS) \cite{NormanSymmetrizedEDCNature}. Note that this procedure is strictly valid at the HS where particle-hole symmetry is satisfied, as indicated by previous model calculations \cite{NormanFermiArcModeling,Zimmers_Millis_EPL_PCCO,PARK_ARPES_SF_PRB2013,TremblayTheorySF} (details in the Supplementary Information). Experimental SEDCs display the filling of the PG at high temperature and allow to extract a PG amplitude $\Delta_{\text{PG}}\approx$85\,meV. The filling of the PG can be well modeled by increasing the spin-fluctuation spectral broadening term $\Gamma$ (from 85\,meV to 160\,meV for 50\,K, black curves, and 130\,K, red curves, respectively), while $\Delta_{\text{PG}}$=$\eta$=85\,meV are fixed. However, we remark that any experimental estimate of the temperature dependence of the PG spectral weight by fitting of SEDCs may be affected by intrinsic and uncorrelated noise. We overcome this limitation by computing the difference between the photoemission intensity for high temperature (130\,K) and its counterpart for low temperature (50\,K), as shown in Fig.\,\ref{Fig1}f. As discussed later in more detail in Eq. \ref{EQ:2}, this differential curve is proportional to the experimental differential momentum-integrated EDCs (dEDCs). It is evident that a filling of the PG may lead to an increase of the photoemission intensity for $\omega \approx$-50\, meV (blue arrows in Fig.\,\ref{Fig1}f).
We note that a pure thermal broadening would suppress (increase) the photoemission intensity symmetrically for all $\omega<$0\, meV ($\omega>$0\, meV), independently of the explored momentum region, within an energy range compatible with the corresponding broadening of the Fermi-Dirac distribution function; however, since at $\omega \approx$-50\,meV thermal contributions are negligible within the range of electronic temperatures explored in this work (50\,meV$\approx$5k$_{\text{B}}$T for T=130\,K), this approach allows to track the evolution of the PG with exquisite sensitivity. 
A similar approach has been recently used to track the electron-boson interaction in hole-doped cuprates via a transient analysis/modeling of the band renormalization (kink) \cite{MillerKink2212}. In the specific case of NCCO a modest kink has been reported (see Ref.\,\cite{Shen_NCCO_arXiv} and Supplementary Information), not affecting our analysis and conclusions. 

\subsection*{Intimate relation of  the pseudogap to the spin-correlation length}
\begin{figure*}
\centering
\includegraphics[scale=0.92]{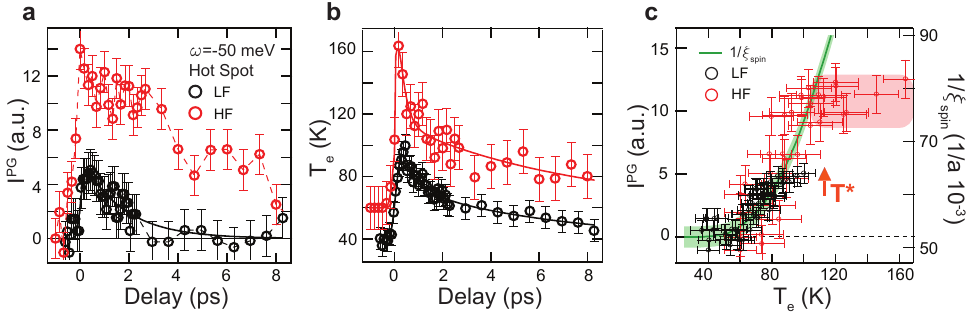}
\caption[Fig2]{\footnotesize Pseudogap spectral weight vs. temperature in optimally-doped NCCO.
\textbf{a}, Temporal evolution of the photoemission intensity at the HS for $\omega$=-50$\pm$10\,meV (see black line and violet dotted circle in Fig.\,\ref{Fig1}b), for both the employed pump fluences LF and HF. The solid black line is a phenomenological single exponential-decay fit for the LF curve while the HF curve saturates in the first 2\,ps after the pump excitation. 
\textbf{b}, Transient electronic temperature extracted by fitting the Fermi edge broadening along the near-nodal direction (green line in Fig.\,\ref{Fig1}b), for both LF and HF. The solid lines are phenomenological double exponential-decay fits.
\textbf{c}, Photoemission intensity at the HS, $\omega$=-50$\pm$10\,meV, as a function of the electronic temperature (black and red circles for LF and HF, respectively). The green line and transparent shadow represent the inverse of the spin-correlation length $\xi_{spin}$ from neutron scattering studies for optimal doping T$_{\text{c}}\approx$24\,K \cite{NeutronNCCO_Nat2007}, appropriately scaled and offset. We identify T* as the temperature at which the PG is completely filled (in agreement with the saturation of the photoemission intensity at the HS for $\omega$=-50$\pm$10\,meV for HF). Error bars in a--c reflect the systematic errors associated with the experiment and the number of averaging cycles acquired for each fluence.}
\label{Fig2} 
\end{figure*}

Having defined the framework for our investigation, we now track the temperature-dependent modification of the low-energy DOS at the HS in optimally-doped NCCO by introducing thermal excitations via optical pumping.
Figure\,\ref{Fig2}a displays the transient enhancement of the photoemission intensity at the HS in a 20\,meV energy window about $\omega$=$-$50\,meV ($I_{\omega=-50}^{PG}$, given by the momentum-integrated EDC along $\varphi\approx$26.5$^o$, momentum direction indicated by the black solid line in Fig.\,\ref{Fig1}b).
This particular choice of energy window was motivated by the dEDCs at the HS (modeled and experimental) shown in Fig.\,\ref{Fig1}f and Fig.\,\ref{Fig3}b.
Two pump fluences were employed, here identified as low fluence (LF) and high fluence (HF).
We note that the temporal response of $I_{\omega=-50}^{PG}$ is dependent on the pump fluence. While the enhancement of $I_{\omega=-50}^{PG}$ recovers exponentially within 2\,ps for the LF regime, in the HF regime $I_{\omega=-50}^{PG}$ saturates for approximately 2\,ps and does not recover within the domain of pump-probe delay studied. \\
In order to establish a direct connection between the TR-ARPES phenomenology and the PG, as well as the reported T*, we must convert the measured time dependence of our data to an effective temperature evolution \cite{NodePeakLanzara}. This is done by fitting the Fermi edge width of the momentum-integrated EDCs along the near-nodal direction ($\varphi\approx$39$^o$, green solid line in Fig.\,\ref{Fig1}b), establishing an effective electronic temperature T$_\text{e}$ for each time delay (Fig.\,\ref{Fig2}b). 
In an effort to unveil the mechanisms which underlie the suppression of the PG spectral weight, we plot $I_{\omega=-50}^{PG}$ directly as a function of T$_{\text{e}}$ in Fig.\,\ref{Fig2}c. We recognize a resemblance between the temperature dependence of $I_{\omega=-50}^{PG}$ and $\xi_{spin}^{-1}$ reported in Ref.\,\cite{NeutronNCCO_Nat2007} for optimal doping (we have superimposed the latter as a green line and shadow in Fig.\,\ref{Fig2}c, with appropriate offset and scaling). However, for temperatures T$_\text{e}>$120\,K, $I_{\omega=-50}^{PG}$(T$_\text{e}$) is found to saturate, marking a departure from $\xi_{spin}^{-1}$. This saturation point is in good agreement with the T* reported by other experimental probes \cite{OpticalPG_NCCO,Matsui_ARPES_PRB2007,OnoseConductivity2004,Zimmers_Millis_EPL_PCCO}. Note that the deviation of $I_{\omega=-50}^{PG}$(T$_\text{e}>$T*) from $\xi_{spin}^{-1}$ is not driven by a phase transition, but is rather a consequence of the PG filling, as we will discuss in the following.  

\section*{Theoretical Model}
\begin{figure*}
\centering
\includegraphics[scale=0.8]{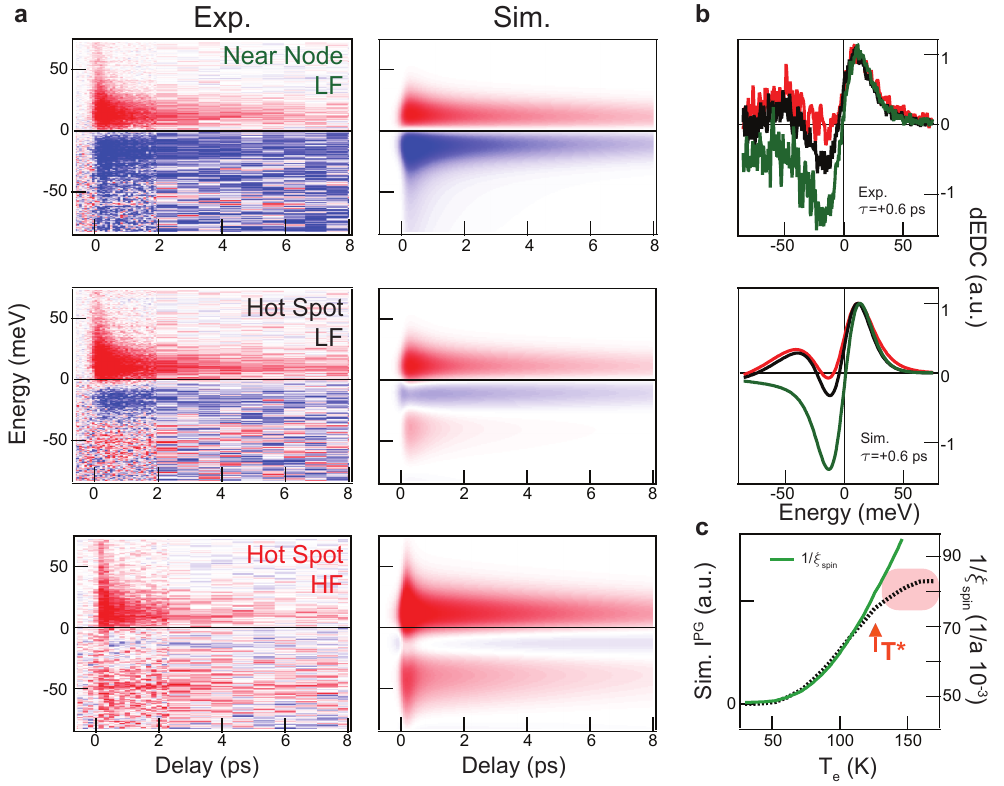}
\caption[Fig3]{\footnotesize Comparison of experimental and simulated TR-ARPES data.
\textbf{a}, Momentum-integrated differential energy distribution curves (dEDCs), experimental (left column) and simulated (right column), along the near-nodal direction (top panels, LF) and at the HS (middle and bottom panels for LF and HF, respectively).
Simulated panels have been generated using the fit of the experimental transient T$_{\text{e}}$ shown in Fig.\,\ref{Fig2}b, and assuming $\Gamma(\text{T}_{\text{e}})= C \cdot \xi_{spin}^{-1}(\text{T}_{\text{e}})$ ($C \approx$1.9\,a\,eV, where a is the unit cell size, details in Supplementary Information), as in Ref.\,\cite{TremblayTheorySF}.
\textbf{b}, Experimental (upper panel) and simulated (bottom panel) dEDCs along the near-nodal direction (green line, LF), and at the HS (black and red lines for LF and HF, respectively), at $\tau$=+0.6\,ps. 
\textbf{c}, Simulated photoemission intensity at the HS, $\omega$=-50$\pm$10\,meV, as a function of the electronic temperature (black dashed line). The green line is the inverse of $\xi_{spin}$, obtained from Ref.\,\cite{NeutronNCCO_Nat2007} as discussed in Fig.\,\ref{Fig2}c.}
\label{Fig3} 
\end{figure*}

To further clarify the origin of T*, we now present a comprehensive analysis and modeling of the photoinduced thermal modification of the PG for both LF and HF pump regimes. The spectral features of NCCO are inherently broad, precluding the sort of detailed analysis of the transient spectral function which has been achieved for hole-doped cuprates \cite{Boschini2018}. Alternately, we focus our analysis on the temporal evolution of dEDCs, defined as:
\begin{equation} \label{EQ:2}
  \text{dEDC}(\omega) \propto \text{DOS}(\omega,\tau)\cdot f(\omega,\tau) - \text{DOS}_0(\omega) \cdot f_0(\omega),
\end{equation}
where $\text{DOS}_0 (\omega)$ and the Fermi-Dirac electronic distribution $f_0(\omega)$ are the unperturbed quantities (see Refs.\,\cite{DessauPRX,MillerKink2212} and Supplementary Information). Note that the photoemission matrix-elements, not included in Eq.\,\ref{EQ:2}, represent a renormalization factor.
The left column of Fig.\,\ref{Fig3}a displays experimental dEDCs as a function of the pump-probe delay ($\tau$) and binding energy ($\omega$) for the near-nodal cut (top panel, LF), and the HS (middle and bottom panels for LF and HF, respectively). 
The TR-ARPES data reported here can be simulated remarkably well using the simple model of Eq.\,\ref{EQ:1} (see right column in Fig.\,\ref{Fig3}a).
The experimental dEDCs are reproduced through a substantial increase of the broadening term $\Gamma$ alone, which phenomenologically describes the loss of quasiparticle intengrity and the ensuing filling of the PG due to the reduction of the spin-correlation length \cite{PARK_ARPES_SF_PRB2013,KyungPRL2004}. 
Note that, $\Gamma(\text{T}_{\text{e}}) \propto \xi_{spin}^{-1}(\text{T}_{\text{e}})$ in Eq.\,\ref{EQ:1} is assumed for simulation purposes, in agreement with theoretical predictions of Vilk and Tremblay \cite{TremblayTheorySF}. A full closure of the gap fails to reproduce our TR-ARPES data (details in the Supplementary Information). 
Instead, the filling of the PG is in agreement with the SEDCs presented in Fig.\,\ref{Fig1}e and scattering studies \cite{XrayNCCO_NatPhys2014,NeutronNCCO_Nat2007}, which show that the spectral weight associated with magnetic excitations displays a much weaker temperature dependence than the one of the spin-correlation length. \\
In further support of our interpretation of the HS data, we compare single experimental and simulated dEDCs along the near-nodal direction and HS for $\tau$=+0.6\,ps in Fig.\,\ref{Fig3}b. Along the near-nodal direction we find an almost symmetric transient population/depletion (increase/decrease of the photoemission intensity for $\omega>$/$<$0 across the Fermi level), characteristic of a mere thermal broadening effect (in the near-nodal region the PG is centered in the unoccupied states, details in the Supplementary Information). In contrast to this, at the HS we observe instead the increment of the intensity for $\omega \approx -$50\,meV and a modest (null) depletion signal for $\omega \approx -$20\,meV for LF (HF), in good agreement with the model and data of Fig.\,\ref{Fig1}d-f and Fig.\,\ref{Fig2}.
Finally, we plot in Fig.\,\ref{Fig3}c the simulated analog to Fig.\,\ref{Fig2}c, noting a remarkable correspondence between the two figures. In particular, assuming the direct relationship between the filling of the PG and $\xi_{spin}$ (as predicted for 2D spin-fluctuations \cite{TremblayTheorySF}), we find that the simulated filling of the PG saturates for temperatures T$\approx$T* when the spin-fluctuations-induced broadening $\Gamma$(T*)$\approx 2 \Delta_{\text{PG}}\approx$170\,meV. We note that this empirical observation agrees well with recent theoretical investigations of the vanishing of the PG in the electron-doped cuprate Pr$_{\text{1.3-x}}$La$_{\text{0.7}}$Ce$_{\text{x}}$CuO$_{\text{4}}$ \cite{TheoryPCCO_fillingPG}. 

\section*{Conclusions}
In conclusion, we have reported the intimate relation between the partial suppression of the electronic spectral weight, a.k.a pseudogap, and the short-range magnetic correlations in electron-doped cuprates.  
In particular, by performing a detailed ultrafast angle-resolved photoemission study at the hot-spot of the optimally-doped NCCO electron-doped cuprate, we have demonstrated that the temperature dependence of the low-energy DOS is closely related to the spin-correlation length $\xi_{spin}$. 
We identified two different temperature regimes for the PG, moving from low to high temperature: i) T$<$T*, in which the PG begins to fill alongside with the reduction of $\xi_{spin}$; ii) T$>$T*, where the PG is completely filled-up. 
Our results show that the PG phenomenology in optimally-doped NCCO originates from short-range AF correlations, parametrized by $\xi_{spin}$(T), and T* is the temperature crossover above which the spin-fluctuation-induced spectral broadening overcomes substantially the PG amplitude $\Delta_{\text{PG}}$, manifest spectroscopically as a loss of quasiparticle integrity and subsequent filling of the PG. Therefore, the frequently reported onset temperature T* does not represent a thermodynamic phase transition, \emph{i.e.} a sudden quenching of a well-defined order parameter; rather, T* is solely driven by the weakening of the short-range AF correlations and incipient ($\pi$,$\pi$)-folding \cite{PARK_ARPES_SF_PRB2013,KyungPRL2004}.
In addition, the filling -- not closure -- of the PG (see Fig.\,\ref{Fig1}e, Fig.\,\ref{Fig2}c, Fig.\,\ref{Fig3}c, and Supplementary Information) suggests that the energy scale associated with the PG survives for temperatures well above T* and bears a relation to the underlying Mott physics: while the PG amplitude is mainly driven by local interactions, PG spectral features are washed out by spin-fluctuations which suppress the quasiparticle integrity.
Finally, we note that our transient momentum-resolved study demonstrates that even underlying orders with correlation lengths of about tens of unit cells may play a significant role in shaping the Fermi surface topology and associated transport properties of complex materials \cite{Shen_NCCO_arXiv,QuantumOscillationsNCCO,TailleferPRX2017,ReviewElecDoped,OpticalPG_NCCO,Matsui_ARPES_PRB2007,OnoseConductivity2004,Armitage_ARPES_PRL2001,Armitage_ARPES_PRL2002,Matsui_ARPES_PRL2005,PARK_ARPES_SF_PRB2013}.   

\section*{Materials and Methods\label{Methods}}
\textbf{Experimental design}\\Our TR-ARPES setup exploits the classic pump-probe scheme. A 6.2 eV probe beam is generated by fourth-harmonic generation of the fundamental wavelength (800 nm, 1.55 eV) of a Ti:sapphire laser (Vitesse Duo and RegA 9000 by Coherent, 250 kHz repetition rate). The 1.55 eV pump beam is split from the source before harmonic generation. The pump and probe beam diameters on the sample are 300 $\mu$m and 150 $\mu$m, respectively. The angle of incidence on the sample is approximately normal for both beams. The two pump incident fluences were 28$\pm$5\,$\mu$J/cm$^2$ for LF and 50$\pm$10\,$\mu$J/cm$^2$ for HF. The pump and probe beams were s-polarized. The sample was cleaved and measured at $<$5$\cdot$10$^{-11}$ torr base pressure and 10\,K temperature. Overall thermal effects have been taken into account and we estimate a temperature of approximately 40\,K and 60\,K for negative pump-probe delays for LF and HF excitations, respectively. Photoemitted electrons were detected by an hemispherical analyzer (Specs Phoibos 150). Energy and temporal resolutions were 17 meV and 250 fs, respectively. 

\textbf{Sample}\\ High quality electron-doped single crystals Nd$_{\text{2-x}}$Ce$_{\text{x}}$CuO$_{\text{4}+\delta}$ were grown by the container-free traveling solvent floating zone (TSFZ) technique. The optimally doped Nd$_{\text{2-x}}$Ce$_{\text{x}}$CuO$_{\text{4}+\delta}$ crystals ($x\approx$0.15) exhibit an oxygen surplus directly after growth which has to be removed by an additional post-growth annealing treatment. Such annealed crystals exhibit a transition temperature T$_{\text{c}}$ of 23.5\,K with transition widths of 1\,K \cite{Erb2010,BookErb}. The very high crystal quality of Nd$_{\text{2-x}}$Ce$_{\text{x}}$CuO$_{\text{4}+\delta}$ is manifest particularly well in magnetic quantum oscillations observed on several samples at different doping levels \cite{QuantumOscillationsNCCO}.

\begin{acknowledgments}
This research was undertaken thanks in part to funding from the Max Planck-UBC-UTokyo Centre for Quantum Materials and the Canada First Research Excellence Fund, Quantum Materials and Future Technologies Program. The work at UBC was supported by the Gordon and Betty Moore Foundation's  EPiQS Initiative, Grant GBMF4779, the Killam, Alfred P. Sloan, and Natural Sciences and Engineering Research Council of Canada's (NSERC's) Steacie Memorial Fellowships (A.D.), the Alexander von Humboldt Fellowship (A.D.), the Canada Research Chairs Program (A.D.), NSERC, Canada Foundation for Innovation (CFI), British Columbia Knowledge Development Fund (BCKDF), and CIFAR Quantum Materials Program. E.R. acknowledges support from the Swiss National Science Foundation (SNSF) grant no. P300P2\_164649.
C.G. acknowledge financial support from MIUR through the PRIN 2015 Programme (Prot. 2015C5SEJJ001) and from Universit\`{a} Cattolica del Sacro Cuore through D.1, D.2.2 and D.3.1 grants.

\textbf{Author contributions}\\
F.B. and M.Z. equally contributed to this work. F.B., M.Z., E.R. and A.D. conceived the investigation. F.B. and M.Z. performed TR-ARPES measurements with the assistance of E.R. and M.M., and F.B., M.Z., E.R., R.P.D., M.M., B.Z., P.N., M.S., S.Z., A.K.M., G.L. were responsible for operation and maintenance of the experimental system. F.B., M.Z., E.R., E.H.d.S.N., C.G., D.J.J. and A.D. were responsible for data analysis and interpretation. A.E. provided NCCO samples. All of the authors discussed the underlying physics and contributed to the manuscript. F.B., M.Z., R.P.D, and A.D. wrote the manuscript. A.D. was responsible for the overall direction, planning and management of the project.

\textbf{Competing interests}\\
The authors declare that they have no competing interests.

\textbf{Data and materials availability}\\
All data needed to evaluate the conclusions in the paper are present in the paper and/or the Supplementary Materials. Additional data may be requested from the authors.
\end{acknowledgments}

\bibliographystyle{apsrev4-1}

\begin{thebibliography}{53}%
\makeatletter
\providecommand \@ifxundefined [1]{%
 \@ifx{#1\undefined}
}%
\providecommand \@ifnum [1]{%
 \ifnum #1\expandafter \@firstoftwo
 \else \expandafter \@secondoftwo
 \fi
}%
\providecommand \@ifx [1]{%
 \ifx #1\expandafter \@firstoftwo
 \else \expandafter \@secondoftwo
 \fi
}%
\providecommand \natexlab [1]{#1}%
\providecommand \enquote  [1]{``#1''}%
\providecommand \bibnamefont  [1]{#1}%
\providecommand \bibfnamefont [1]{#1}%
\providecommand \citenamefont [1]{#1}%
\providecommand \href@noop [0]{\@secondoftwo}%
\providecommand \href [0]{\begingroup \@sanitize@url \@href}%
\providecommand \@href[1]{\@@startlink{#1}\@@href}%
\providecommand \@@href[1]{\endgroup#1\@@endlink}%
\providecommand \@sanitize@url [0]{\catcode `\\12\catcode `\$12\catcode
  `\&12\catcode `\#12\catcode `\^12\catcode `\_12\catcode `\%12\relax}%
\providecommand \@@startlink[1]{}%
\providecommand \@@endlink[0]{}%
\providecommand \url  [0]{\begingroup\@sanitize@url \@url }%
\providecommand \@url [1]{\endgroup\@href {#1}{\urlprefix }}%
\providecommand \urlprefix  [0]{URL }%
\providecommand \Eprint [0]{\href }%
\providecommand \doibase [0]{http://dx.doi.org/}%
\providecommand \selectlanguage [0]{\@gobble}%
\providecommand \bibinfo  [0]{\@secondoftwo}%
\providecommand \bibfield  [0]{\@secondoftwo}%
\providecommand \translation [1]{[#1]}%
\providecommand \BibitemOpen [0]{}%
\providecommand \bibitemStop [0]{}%
\providecommand \bibitemNoStop [0]{.\EOS\space}%
\providecommand \EOS [0]{\spacefactor3000\relax}%
\providecommand \BibitemShut  [1]{\csname bibitem#1\endcsname}%
\let\auto@bib@innerbib\@empty
\bibitem [{\citenamefont {Fradkin}\ \emph {et~al.}(2015)\citenamefont
  {Fradkin}, \citenamefont {Kivelson},\ and\ \citenamefont
  {Tranquada}}]{FradkinReviewCupratesOrders}%
  \BibitemOpen
  \bibfield  {author} {\bibinfo {author} {\bibfnamefont {E.}~\bibnamefont
  {Fradkin}}, \bibinfo {author} {\bibfnamefont {S.~A.}\ \bibnamefont
  {Kivelson}}, \ and\ \bibinfo {author} {\bibfnamefont {J.~M.}\ \bibnamefont
  {Tranquada}},\ }\href {\doibase 10.1103/RevModPhys.87.457} {\bibfield
  {journal} {\bibinfo  {journal} {Rev. Mod. Phys.}\ }\textbf {\bibinfo {volume}
  {87}},\ \bibinfo {pages} {457} (\bibinfo {year} {2015})}\BibitemShut
  {NoStop}%
\bibitem [{\citenamefont {Keimer}\ and\ \citenamefont
  {Moore}(2017)}]{ReviewKeimerQuantumMat}%
  \BibitemOpen
  \bibfield  {author} {\bibinfo {author} {\bibfnamefont {B.}~\bibnamefont
  {Keimer}}\ and\ \bibinfo {author} {\bibfnamefont {J.~E.}\ \bibnamefont
  {Moore}},\ }\href {\doibase 10.1038/nphys4302} {\bibfield  {journal}
  {\bibinfo  {journal} {Nature Physics}\ }\textbf {\bibinfo {volume} {13}},\
  \bibinfo {pages} {1045} (\bibinfo {year} {2017})}\BibitemShut {NoStop}%
\bibitem [{\citenamefont {Basov}\ \emph {et~al.}(2017)\citenamefont {Basov},
  \citenamefont {Averitt},\ and\ \citenamefont {Hsieh}}]{ReviewBasov}%
  \BibitemOpen
  \bibfield  {author} {\bibinfo {author} {\bibfnamefont {D.~N.}\ \bibnamefont
  {Basov}}, \bibinfo {author} {\bibfnamefont {R.~D.}\ \bibnamefont {Averitt}},
  \ and\ \bibinfo {author} {\bibfnamefont {D.}~\bibnamefont {Hsieh}},\ }\href
  {\doibase 10.1038/nmat5017} {\bibfield  {journal} {\bibinfo  {journal}
  {Nature Materials}\ }\textbf {\bibinfo {volume} {16}},\ \bibinfo {pages}
  {1077} (\bibinfo {year} {2017})}\BibitemShut {NoStop}%
\bibitem [{\citenamefont {Davis}\ and\ \citenamefont
  {Lee}(2013)}]{PNAS_Davis_2013}%
  \BibitemOpen
  \bibfield  {author} {\bibinfo {author} {\bibfnamefont {J.~C.~S.}\
  \bibnamefont {Davis}}\ and\ \bibinfo {author} {\bibfnamefont {D.-H.}\
  \bibnamefont {Lee}},\ }\href {\doibase 10.1073/pnas.1316512110} {\bibfield
  {journal} {\bibinfo  {journal} {Proceedings of the National Academy of
  Sciences}\ }\textbf {\bibinfo {volume} {110}},\ \bibinfo {pages} {17623}
  (\bibinfo {year} {2013})}\BibitemShut {NoStop}%
\bibitem [{\citenamefont {Timusk}\ and\ \citenamefont
  {Statt}(1999)}]{Timusk_RepProgPhys_1999}%
  \BibitemOpen
  \bibfield  {author} {\bibinfo {author} {\bibfnamefont {T.}~\bibnamefont
  {Timusk}}\ and\ \bibinfo {author} {\bibfnamefont {B.}~\bibnamefont {Statt}},\
  }\href@noop {} {\bibfield  {journal} {\bibinfo  {journal} {Reports on
  Progress in Physics}\ }\textbf {\bibinfo {volume} {62}},\ \bibinfo {pages}
  {61} (\bibinfo {year} {1999})}\BibitemShut {NoStop}%
\bibitem [{\citenamefont {H\"{u}fner}\ \emph {et~al.}(2008)\citenamefont
  {H\"{u}fner}, \citenamefont {Hossain}, \citenamefont {Damascelli},\ and\
  \citenamefont {Sawatzky}}]{Hufner_RepProgPhys_2008}%
  \BibitemOpen
  \bibfield  {author} {\bibinfo {author} {\bibfnamefont {S.}~\bibnamefont
  {H\"{u}fner}}, \bibinfo {author} {\bibfnamefont {M.~A.}\ \bibnamefont
  {Hossain}}, \bibinfo {author} {\bibfnamefont {A.}~\bibnamefont {Damascelli}},
  \ and\ \bibinfo {author} {\bibfnamefont {G.~A.}\ \bibnamefont {Sawatzky}},\
  }\href@noop {} {\bibfield  {journal} {\bibinfo  {journal} {Reports on
  Progress in Physics}\ }\textbf {\bibinfo {volume} {71}},\ \bibinfo {pages}
  {062501} (\bibinfo {year} {2008})}\BibitemShut {NoStop}%
\bibitem [{\citenamefont {Vishik}(2018)}]{Vishik_RepProgPhys_2018}%
  \BibitemOpen
  \bibfield  {author} {\bibinfo {author} {\bibfnamefont {I.~M.}\ \bibnamefont
  {Vishik}},\ }\href@noop {} {\bibfield  {journal} {\bibinfo  {journal}
  {Reports on Progress in Physics}\ }\textbf {\bibinfo {volume} {81}},\
  \bibinfo {pages} {062501} (\bibinfo {year} {2018})}\BibitemShut {NoStop}%
\bibitem [{\citenamefont {Borisenko}\ \emph {et~al.}(2008)\citenamefont
  {Borisenko}, \citenamefont {Kordyuk}, \citenamefont {Yaresko}, \citenamefont
  {Zabolotnyy}, \citenamefont {Inosov}, \citenamefont {Schuster}, \citenamefont
  {B\"uchner}, \citenamefont {Weber}, \citenamefont {Follath}, \citenamefont
  {Patthey},\ and\ \citenamefont {Berger}}]{PseudogapCDWBorisenko}%
  \BibitemOpen
  \bibfield  {author} {\bibinfo {author} {\bibfnamefont {S.~V.}\ \bibnamefont
  {Borisenko}}, \bibinfo {author} {\bibfnamefont {A.~A.}\ \bibnamefont
  {Kordyuk}}, \bibinfo {author} {\bibfnamefont {A.~N.}\ \bibnamefont
  {Yaresko}}, \bibinfo {author} {\bibfnamefont {V.~B.}\ \bibnamefont
  {Zabolotnyy}}, \bibinfo {author} {\bibfnamefont {D.~S.}\ \bibnamefont
  {Inosov}}, \bibinfo {author} {\bibfnamefont {R.}~\bibnamefont {Schuster}},
  \bibinfo {author} {\bibfnamefont {B.}~\bibnamefont {B\"uchner}}, \bibinfo
  {author} {\bibfnamefont {R.}~\bibnamefont {Weber}}, \bibinfo {author}
  {\bibfnamefont {R.}~\bibnamefont {Follath}}, \bibinfo {author} {\bibfnamefont
  {L.}~\bibnamefont {Patthey}}, \ and\ \bibinfo {author} {\bibfnamefont
  {H.}~\bibnamefont {Berger}},\ }\href {\doibase
  10.1103/PhysRevLett.100.196402} {\bibfield  {journal} {\bibinfo  {journal}
  {Phys. Rev. Lett.}\ }\textbf {\bibinfo {volume} {100}},\ \bibinfo {pages}
  {196402} (\bibinfo {year} {2008})}\BibitemShut {NoStop}%
\bibitem [{\citenamefont {Rossnagel}(2011)}]{ReviewRossnagel}%
  \BibitemOpen
  \bibfield  {author} {\bibinfo {author} {\bibfnamefont {K.}~\bibnamefont
  {Rossnagel}},\ }\href {http://stacks.iop.org/0953-8984/23/i=21/a=213001}
  {\bibfield  {journal} {\bibinfo  {journal} {Journal of Physics: Condensed
  Matter}\ }\textbf {\bibinfo {volume} {23}},\ \bibinfo {pages} {213001}
  (\bibinfo {year} {2011})}\BibitemShut {NoStop}%
\bibitem [{\citenamefont {Stajic}\ \emph {et~al.}(2004)\citenamefont {Stajic},
  \citenamefont {Milstein}, \citenamefont {Chen}, \citenamefont {Chiofalo},
  \citenamefont {Holland},\ and\ \citenamefont {Levin}}]{PseudogapColdAtoms}%
  \BibitemOpen
  \bibfield  {author} {\bibinfo {author} {\bibfnamefont {J.}~\bibnamefont
  {Stajic}}, \bibinfo {author} {\bibfnamefont {J.~N.}\ \bibnamefont
  {Milstein}}, \bibinfo {author} {\bibfnamefont {Q.}~\bibnamefont {Chen}},
  \bibinfo {author} {\bibfnamefont {M.~L.}\ \bibnamefont {Chiofalo}}, \bibinfo
  {author} {\bibfnamefont {M.~J.}\ \bibnamefont {Holland}}, \ and\ \bibinfo
  {author} {\bibfnamefont {K.}~\bibnamefont {Levin}},\ }\href {\doibase
  10.1103/PhysRevA.69.063610} {\bibfield  {journal} {\bibinfo  {journal} {Phys.
  Rev. A}\ }\textbf {\bibinfo {volume} {69}},\ \bibinfo {pages} {063610}
  (\bibinfo {year} {2004})}\BibitemShut {NoStop}%
\bibitem [{\citenamefont {Randeria}(2010)}]{RanderiaColdAtoms}%
  \BibitemOpen
  \bibfield  {author} {\bibinfo {author} {\bibfnamefont {M.}~\bibnamefont
  {Randeria}},\ }\href {\doibase 10.1038/nphys1748} {\bibfield  {journal}
  {\bibinfo  {journal} {Nature Physics}\ }\textbf {\bibinfo {volume} {6}},\
  \bibinfo {pages} {561} (\bibinfo {year} {2010})}\BibitemShut {NoStop}%
\bibitem [{\citenamefont {Chen}\ and\ \citenamefont
  {Wang}(2014)}]{Chen_ReviewPGcoldAtoms_2014}%
  \BibitemOpen
  \bibfield  {author} {\bibinfo {author} {\bibfnamefont {Q.}~\bibnamefont
  {Chen}}\ and\ \bibinfo {author} {\bibfnamefont {J.}~\bibnamefont {Wang}},\
  }\href {\doibase 10.1007/s11467-014-0448-7} {\bibfield  {journal} {\bibinfo
  {journal} {Frontiers of Physics}\ }\textbf {\bibinfo {volume} {9}},\ \bibinfo
  {pages} {539} (\bibinfo {year} {2014})}\BibitemShut {NoStop}%
\bibitem [{\citenamefont {Mott}(1968)}]{ReviewMott_MIT}%
  \BibitemOpen
  \bibfield  {author} {\bibinfo {author} {\bibfnamefont {N.~F.}\ \bibnamefont
  {Mott}},\ }\href {\doibase 10.1103/RevModPhys.40.677} {\bibfield  {journal}
  {\bibinfo  {journal} {Rev. Mod. Phys.}\ }\textbf {\bibinfo {volume} {40}},\
  \bibinfo {pages} {677} (\bibinfo {year} {1968})}\BibitemShut {NoStop}%
\bibitem [{\citenamefont {Kordyuk}(2015)}]{ReviewKordyuk}%
  \BibitemOpen
  \bibfield  {author} {\bibinfo {author} {\bibfnamefont {A.~A.}\ \bibnamefont
  {Kordyuk}},\ }\href {\doibase 10.1063/1.4919371} {\bibfield  {journal}
  {\bibinfo  {journal} {Low Temperature Physics}\ }\textbf {\bibinfo {volume}
  {41}},\ \bibinfo {pages} {319} (\bibinfo {year} {2015})}\BibitemShut
  {NoStop}%
\bibitem [{\citenamefont {Norman}\ \emph {et~al.}(2007)\citenamefont {Norman},
  \citenamefont {Kanigel}, \citenamefont {Randeria}, \citenamefont
  {Chatterjee},\ and\ \citenamefont {Campuzano}}]{NormanFermiArcModeling}%
  \BibitemOpen
  \bibfield  {author} {\bibinfo {author} {\bibfnamefont {M.~R.}\ \bibnamefont
  {Norman}}, \bibinfo {author} {\bibfnamefont {A.}~\bibnamefont {Kanigel}},
  \bibinfo {author} {\bibfnamefont {M.}~\bibnamefont {Randeria}}, \bibinfo
  {author} {\bibfnamefont {U.}~\bibnamefont {Chatterjee}}, \ and\ \bibinfo
  {author} {\bibfnamefont {J.~C.}\ \bibnamefont {Campuzano}},\ }\href {\doibase
  10.1103/PhysRevB.76.174501} {\bibfield  {journal} {\bibinfo  {journal} {Phys.
  Rev. B}\ }\textbf {\bibinfo {volume} {76}},\ \bibinfo {pages} {174501}
  (\bibinfo {year} {2007})}\BibitemShut {NoStop}%
\bibitem [{\citenamefont {Keimer}\ \emph {et~al.}(2015)\citenamefont {Keimer},
  \citenamefont {Kivelson}, \citenamefont {Norman}, \citenamefont {Uchida},\
  and\ \citenamefont {Zaanen}}]{KeimerReviewNature}%
  \BibitemOpen
  \bibfield  {author} {\bibinfo {author} {\bibfnamefont {B.}~\bibnamefont
  {Keimer}}, \bibinfo {author} {\bibfnamefont {S.~A.}\ \bibnamefont
  {Kivelson}}, \bibinfo {author} {\bibfnamefont {M.~R.}\ \bibnamefont
  {Norman}}, \bibinfo {author} {\bibfnamefont {S.}~\bibnamefont {Uchida}}, \
  and\ \bibinfo {author} {\bibfnamefont {J.}~\bibnamefont {Zaanen}},\ }\href
  {\doibase 10.1038/nature14165} {\bibfield  {journal} {\bibinfo  {journal}
  {Nature}\ }\textbf {\bibinfo {volume} {518}},\ \bibinfo {pages} {179}
  (\bibinfo {year} {2015})}\BibitemShut {NoStop}%
\bibitem [{\citenamefont {Sato}\ \emph {et~al.}(2017)\citenamefont {Sato},
  \citenamefont {Kasahara}, \citenamefont {Murayama}, \citenamefont {Kasahara},
  \citenamefont {Moon}, \citenamefont {Nishizaki}, \citenamefont {Loew},
  \citenamefont {Porras}, \citenamefont {Keimer}, \citenamefont {Shibauchi},\
  and\ \citenamefont {Matsuda}}]{ThermodynYBCONematic}%
  \BibitemOpen
  \bibfield  {author} {\bibinfo {author} {\bibfnamefont {Y.}~\bibnamefont
  {Sato}}, \bibinfo {author} {\bibfnamefont {S.}~\bibnamefont {Kasahara}},
  \bibinfo {author} {\bibfnamefont {H.}~\bibnamefont {Murayama}}, \bibinfo
  {author} {\bibfnamefont {Y.}~\bibnamefont {Kasahara}}, \bibinfo {author}
  {\bibfnamefont {E.-G.}\ \bibnamefont {Moon}}, \bibinfo {author}
  {\bibfnamefont {T.}~\bibnamefont {Nishizaki}}, \bibinfo {author}
  {\bibfnamefont {T.}~\bibnamefont {Loew}}, \bibinfo {author} {\bibfnamefont
  {J.}~\bibnamefont {Porras}}, \bibinfo {author} {\bibfnamefont
  {B.}~\bibnamefont {Keimer}}, \bibinfo {author} {\bibfnamefont
  {T.}~\bibnamefont {Shibauchi}}, \ and\ \bibinfo {author} {\bibfnamefont
  {Y.}~\bibnamefont {Matsuda}},\ }\href {\doibase 10.1038/nphys4205} {\bibfield
   {journal} {\bibinfo  {journal} {Nature Physics}\ }\textbf {\bibinfo {volume}
  {13}},\ \bibinfo {pages} {1074} (\bibinfo {year} {2017})}\BibitemShut
  {NoStop}%
\bibitem [{\citenamefont {Armitage}\ \emph {et~al.}(2010)\citenamefont
  {Armitage}, \citenamefont {Fournier},\ and\ \citenamefont
  {Greene}}]{ReviewElecDoped}%
  \BibitemOpen
  \bibfield  {author} {\bibinfo {author} {\bibfnamefont {N.~P.}\ \bibnamefont
  {Armitage}}, \bibinfo {author} {\bibfnamefont {P.}~\bibnamefont {Fournier}},
  \ and\ \bibinfo {author} {\bibfnamefont {R.~L.}\ \bibnamefont {Greene}},\
  }\href {\doibase 10.1103/RevModPhys.82.2421} {\bibfield  {journal} {\bibinfo
  {journal} {Rev. Mod. Phys.}\ }\textbf {\bibinfo {volume} {82}},\ \bibinfo
  {pages} {2421} (\bibinfo {year} {2010})}\BibitemShut {NoStop}%
\bibitem [{\citenamefont {Motoyama}\ \emph {et~al.}(2007)\citenamefont
  {Motoyama}, \citenamefont {Yu}, \citenamefont {Vishik}, \citenamefont {Vajk},
  \citenamefont {Mang},\ and\ \citenamefont {Greven}}]{NeutronNCCO_Nat2007}%
  \BibitemOpen
  \bibfield  {author} {\bibinfo {author} {\bibfnamefont {E.~M.}\ \bibnamefont
  {Motoyama}}, \bibinfo {author} {\bibfnamefont {G.}~\bibnamefont {Yu}},
  \bibinfo {author} {\bibfnamefont {I.~M.}\ \bibnamefont {Vishik}}, \bibinfo
  {author} {\bibfnamefont {O.~P.}\ \bibnamefont {Vajk}}, \bibinfo {author}
  {\bibfnamefont {P.~K.}\ \bibnamefont {Mang}}, \ and\ \bibinfo {author}
  {\bibfnamefont {M.}~\bibnamefont {Greven}},\ }\href {\doibase
  10.1038/nature05437} {\bibfield  {journal} {\bibinfo  {journal} {Nature}\
  }\textbf {\bibinfo {volume} {445}},\ \bibinfo {pages} {186} (\bibinfo {year}
  {2007})}\BibitemShut {NoStop}%
\bibitem [{\citenamefont {Onose}\ \emph {et~al.}(2001)\citenamefont {Onose},
  \citenamefont {Taguchi}, \citenamefont {Ishizaka},\ and\ \citenamefont
  {Tokura}}]{OpticalPG_NCCO}%
  \BibitemOpen
  \bibfield  {author} {\bibinfo {author} {\bibfnamefont {Y.}~\bibnamefont
  {Onose}}, \bibinfo {author} {\bibfnamefont {Y.}~\bibnamefont {Taguchi}},
  \bibinfo {author} {\bibfnamefont {K.}~\bibnamefont {Ishizaka}}, \ and\
  \bibinfo {author} {\bibfnamefont {Y.}~\bibnamefont {Tokura}},\ }\href
  {\doibase 10.1103/PhysRevLett.87.217001} {\bibfield  {journal} {\bibinfo
  {journal} {Phys. Rev. Lett.}\ }\textbf {\bibinfo {volume} {87}},\ \bibinfo
  {pages} {217001} (\bibinfo {year} {2001})}\BibitemShut {NoStop}%
\bibitem [{\citenamefont {Matsui}\ \emph {et~al.}(2007)\citenamefont {Matsui},
  \citenamefont {Takahashi}, \citenamefont {Sato}, \citenamefont {Terashima},
  \citenamefont {Ding}, \citenamefont {Uefuji},\ and\ \citenamefont
  {Yamada}}]{Matsui_ARPES_PRB2007}%
  \BibitemOpen
  \bibfield  {author} {\bibinfo {author} {\bibfnamefont {H.}~\bibnamefont
  {Matsui}}, \bibinfo {author} {\bibfnamefont {T.}~\bibnamefont {Takahashi}},
  \bibinfo {author} {\bibfnamefont {T.}~\bibnamefont {Sato}}, \bibinfo {author}
  {\bibfnamefont {K.}~\bibnamefont {Terashima}}, \bibinfo {author}
  {\bibfnamefont {H.}~\bibnamefont {Ding}}, \bibinfo {author} {\bibfnamefont
  {T.}~\bibnamefont {Uefuji}}, \ and\ \bibinfo {author} {\bibfnamefont
  {K.}~\bibnamefont {Yamada}},\ }\href {\doibase 10.1103/PhysRevB.75.224514}
  {\bibfield  {journal} {\bibinfo  {journal} {Phys. Rev. B}\ }\textbf {\bibinfo
  {volume} {75}},\ \bibinfo {pages} {224514} (\bibinfo {year}
  {2007})}\BibitemShut {NoStop}%
\bibitem [{\citenamefont {Onose}\ \emph {et~al.}(2004)\citenamefont {Onose},
  \citenamefont {Taguchi}, \citenamefont {Ishizaka},\ and\ \citenamefont
  {Tokura}}]{OnoseConductivity2004}%
  \BibitemOpen
  \bibfield  {author} {\bibinfo {author} {\bibfnamefont {Y.}~\bibnamefont
  {Onose}}, \bibinfo {author} {\bibfnamefont {Y.}~\bibnamefont {Taguchi}},
  \bibinfo {author} {\bibfnamefont {K.}~\bibnamefont {Ishizaka}}, \ and\
  \bibinfo {author} {\bibfnamefont {Y.}~\bibnamefont {Tokura}},\ }\href
  {\doibase 10.1103/PhysRevB.69.024504} {\bibfield  {journal} {\bibinfo
  {journal} {Phys. Rev. B}\ }\textbf {\bibinfo {volume} {69}},\ \bibinfo
  {pages} {024504} (\bibinfo {year} {2004})}\BibitemShut {NoStop}%
\bibitem [{\citenamefont {Armitage}\ \emph {et~al.}(2001)\citenamefont
  {Armitage}, \citenamefont {Lu}, \citenamefont {Kim}, \citenamefont
  {Damascelli}, \citenamefont {Shen}, \citenamefont {Ronning}, \citenamefont
  {Feng}, \citenamefont {Bogdanov}, \citenamefont {Shen}, \citenamefont
  {Onose}, \citenamefont {Taguchi}, \citenamefont {Tokura}, \citenamefont
  {Mang}, \citenamefont {Kaneko},\ and\ \citenamefont
  {Greven}}]{Armitage_ARPES_PRL2001}%
  \BibitemOpen
  \bibfield  {author} {\bibinfo {author} {\bibfnamefont {N.~P.}\ \bibnamefont
  {Armitage}}, \bibinfo {author} {\bibfnamefont {D.~H.}\ \bibnamefont {Lu}},
  \bibinfo {author} {\bibfnamefont {C.}~\bibnamefont {Kim}}, \bibinfo {author}
  {\bibfnamefont {A.}~\bibnamefont {Damascelli}}, \bibinfo {author}
  {\bibfnamefont {K.~M.}\ \bibnamefont {Shen}}, \bibinfo {author}
  {\bibfnamefont {F.}~\bibnamefont {Ronning}}, \bibinfo {author} {\bibfnamefont
  {D.~L.}\ \bibnamefont {Feng}}, \bibinfo {author} {\bibfnamefont
  {P.}~\bibnamefont {Bogdanov}}, \bibinfo {author} {\bibfnamefont {Z.-X.}\
  \bibnamefont {Shen}}, \bibinfo {author} {\bibfnamefont {Y.}~\bibnamefont
  {Onose}}, \bibinfo {author} {\bibfnamefont {Y.}~\bibnamefont {Taguchi}},
  \bibinfo {author} {\bibfnamefont {Y.}~\bibnamefont {Tokura}}, \bibinfo
  {author} {\bibfnamefont {P.~K.}\ \bibnamefont {Mang}}, \bibinfo {author}
  {\bibfnamefont {N.}~\bibnamefont {Kaneko}}, \ and\ \bibinfo {author}
  {\bibfnamefont {M.}~\bibnamefont {Greven}},\ }\href {\doibase
  10.1103/PhysRevLett.87.147003} {\bibfield  {journal} {\bibinfo  {journal}
  {Phys. Rev. Lett.}\ }\textbf {\bibinfo {volume} {87}},\ \bibinfo {pages}
  {147003} (\bibinfo {year} {2001})}\BibitemShut {NoStop}%
\bibitem [{\citenamefont {Armitage}\ \emph {et~al.}(2002)\citenamefont
  {Armitage}, \citenamefont {Ronning}, \citenamefont {Lu}, \citenamefont {Kim},
  \citenamefont {Damascelli}, \citenamefont {Shen}, \citenamefont {Feng},
  \citenamefont {Eisaki}, \citenamefont {Shen}, \citenamefont {Mang},
  \citenamefont {Kaneko}, \citenamefont {Greven}, \citenamefont {Onose},
  \citenamefont {Taguchi},\ and\ \citenamefont
  {Tokura}}]{Armitage_ARPES_PRL2002}%
  \BibitemOpen
  \bibfield  {author} {\bibinfo {author} {\bibfnamefont {N.~P.}\ \bibnamefont
  {Armitage}}, \bibinfo {author} {\bibfnamefont {F.}~\bibnamefont {Ronning}},
  \bibinfo {author} {\bibfnamefont {D.~H.}\ \bibnamefont {Lu}}, \bibinfo
  {author} {\bibfnamefont {C.}~\bibnamefont {Kim}}, \bibinfo {author}
  {\bibfnamefont {A.}~\bibnamefont {Damascelli}}, \bibinfo {author}
  {\bibfnamefont {K.~M.}\ \bibnamefont {Shen}}, \bibinfo {author}
  {\bibfnamefont {D.~L.}\ \bibnamefont {Feng}}, \bibinfo {author}
  {\bibfnamefont {H.}~\bibnamefont {Eisaki}}, \bibinfo {author} {\bibfnamefont
  {Z.-X.}\ \bibnamefont {Shen}}, \bibinfo {author} {\bibfnamefont {P.~K.}\
  \bibnamefont {Mang}}, \bibinfo {author} {\bibfnamefont {N.}~\bibnamefont
  {Kaneko}}, \bibinfo {author} {\bibfnamefont {M.}~\bibnamefont {Greven}},
  \bibinfo {author} {\bibfnamefont {Y.}~\bibnamefont {Onose}}, \bibinfo
  {author} {\bibfnamefont {Y.}~\bibnamefont {Taguchi}}, \ and\ \bibinfo
  {author} {\bibfnamefont {Y.}~\bibnamefont {Tokura}},\ }\href {\doibase
  10.1103/PhysRevLett.88.257001} {\bibfield  {journal} {\bibinfo  {journal}
  {Phys. Rev. Lett.}\ }\textbf {\bibinfo {volume} {88}},\ \bibinfo {pages}
  {257001} (\bibinfo {year} {2002})}\BibitemShut {NoStop}%
\bibitem [{\citenamefont {Matsui}\ \emph {et~al.}(2005)\citenamefont {Matsui},
  \citenamefont {Terashima}, \citenamefont {Sato}, \citenamefont {Takahashi},
  \citenamefont {Wang}, \citenamefont {Yang}, \citenamefont {Ding},
  \citenamefont {Uefuji},\ and\ \citenamefont {Yamada}}]{Matsui_ARPES_PRL2005}%
  \BibitemOpen
  \bibfield  {author} {\bibinfo {author} {\bibfnamefont {H.}~\bibnamefont
  {Matsui}}, \bibinfo {author} {\bibfnamefont {K.}~\bibnamefont {Terashima}},
  \bibinfo {author} {\bibfnamefont {T.}~\bibnamefont {Sato}}, \bibinfo {author}
  {\bibfnamefont {T.}~\bibnamefont {Takahashi}}, \bibinfo {author}
  {\bibfnamefont {S.-C.}\ \bibnamefont {Wang}}, \bibinfo {author}
  {\bibfnamefont {H.-B.}\ \bibnamefont {Yang}}, \bibinfo {author}
  {\bibfnamefont {H.}~\bibnamefont {Ding}}, \bibinfo {author} {\bibfnamefont
  {T.}~\bibnamefont {Uefuji}}, \ and\ \bibinfo {author} {\bibfnamefont
  {K.}~\bibnamefont {Yamada}},\ }\href {\doibase 10.1103/PhysRevLett.94.047005}
  {\bibfield  {journal} {\bibinfo  {journal} {Phys. Rev. Lett.}\ }\textbf
  {\bibinfo {volume} {94}},\ \bibinfo {pages} {047005} (\bibinfo {year}
  {2005})}\BibitemShut {NoStop}%
\bibitem [{\citenamefont {Park}\ \emph {et~al.}(2013)\citenamefont {Park},
  \citenamefont {Morinari}, \citenamefont {Song}, \citenamefont {Leem},
  \citenamefont {Kim}, \citenamefont {Choi}, \citenamefont {Choi},
  \citenamefont {Kim}, \citenamefont {Schmitt}, \citenamefont {Mo},
  \citenamefont {Lu}, \citenamefont {Shen}, \citenamefont {Eisaki},
  \citenamefont {Tohyama}, \citenamefont {Han},\ and\ \citenamefont
  {Kim}}]{PARK_ARPES_SF_PRB2013}%
  \BibitemOpen
  \bibfield  {author} {\bibinfo {author} {\bibfnamefont {S.~R.}\ \bibnamefont
  {Park}}, \bibinfo {author} {\bibfnamefont {T.}~\bibnamefont {Morinari}},
  \bibinfo {author} {\bibfnamefont {D.~J.}\ \bibnamefont {Song}}, \bibinfo
  {author} {\bibfnamefont {C.~S.}\ \bibnamefont {Leem}}, \bibinfo {author}
  {\bibfnamefont {C.}~\bibnamefont {Kim}}, \bibinfo {author} {\bibfnamefont
  {S.~K.}\ \bibnamefont {Choi}}, \bibinfo {author} {\bibfnamefont
  {K.}~\bibnamefont {Choi}}, \bibinfo {author} {\bibfnamefont {J.~H.}\
  \bibnamefont {Kim}}, \bibinfo {author} {\bibfnamefont {F.}~\bibnamefont
  {Schmitt}}, \bibinfo {author} {\bibfnamefont {S.~K.}\ \bibnamefont {Mo}},
  \bibinfo {author} {\bibfnamefont {D.~H.}\ \bibnamefont {Lu}}, \bibinfo
  {author} {\bibfnamefont {Z.-X.}\ \bibnamefont {Shen}}, \bibinfo {author}
  {\bibfnamefont {H.}~\bibnamefont {Eisaki}}, \bibinfo {author} {\bibfnamefont
  {T.}~\bibnamefont {Tohyama}}, \bibinfo {author} {\bibfnamefont {J.~H.}\
  \bibnamefont {Han}}, \ and\ \bibinfo {author} {\bibfnamefont
  {C.}~\bibnamefont {Kim}},\ }\href {\doibase 10.1103/PhysRevB.87.174527}
  {\bibfield  {journal} {\bibinfo  {journal} {Phys. Rev. B}\ }\textbf {\bibinfo
  {volume} {87}},\ \bibinfo {pages} {174527} (\bibinfo {year}
  {2013})}\BibitemShut {NoStop}%
\bibitem [{\citenamefont {Zimmers}\ \emph {et~al.}(2005)\citenamefont
  {Zimmers}, \citenamefont {Tomczak}, \citenamefont {Lobo}, \citenamefont
  {Bontemps}, \citenamefont {Hill}, \citenamefont {Barr}, \citenamefont
  {Dagan}, \citenamefont {Greene}, \citenamefont {Millis},\ and\ \citenamefont
  {Homes}}]{Zimmers_Millis_EPL_PCCO}%
  \BibitemOpen
  \bibfield  {author} {\bibinfo {author} {\bibfnamefont {A.}~\bibnamefont
  {Zimmers}}, \bibinfo {author} {\bibfnamefont {J.~M.}\ \bibnamefont
  {Tomczak}}, \bibinfo {author} {\bibfnamefont {R.~P. S.~M.}\ \bibnamefont
  {Lobo}}, \bibinfo {author} {\bibfnamefont {N.}~\bibnamefont {Bontemps}},
  \bibinfo {author} {\bibfnamefont {C.~P.}\ \bibnamefont {Hill}}, \bibinfo
  {author} {\bibfnamefont {M.~C.}\ \bibnamefont {Barr}}, \bibinfo {author}
  {\bibfnamefont {Y.}~\bibnamefont {Dagan}}, \bibinfo {author} {\bibfnamefont
  {R.~L.}\ \bibnamefont {Greene}}, \bibinfo {author} {\bibfnamefont {A.~J.}\
  \bibnamefont {Millis}}, \ and\ \bibinfo {author} {\bibfnamefont {C.~C.}\
  \bibnamefont {Homes}},\ }\href {http://stacks.iop.org/0295-5075/70/i=2/a=225}
  {\bibfield  {journal} {\bibinfo  {journal} {EPL (Europhysics Letters)}\
  }\textbf {\bibinfo {volume} {70}},\ \bibinfo {pages} {225} (\bibinfo {year}
  {2005})}\BibitemShut {NoStop}%
\bibitem [{\citenamefont {Lee}\ \emph {et~al.}(2014)\citenamefont {Lee},
  \citenamefont {Lee}, \citenamefont {Nowadnick}, \citenamefont {Gerber},
  \citenamefont {Tabis}, \citenamefont {Huang}, \citenamefont {Strocov},
  \citenamefont {Motoyama}, \citenamefont {Yu}, \citenamefont {Moritz},
  \citenamefont {Huang}, \citenamefont {Wang}, \citenamefont {Huang},
  \citenamefont {Wu}, \citenamefont {Chen}, \citenamefont {Huang},
  \citenamefont {Greven}, \citenamefont {Schmitt}, \citenamefont {Shen},\ and\
  \citenamefont {Devereaux}}]{XrayNCCO_NatPhys2014}%
  \BibitemOpen
  \bibfield  {author} {\bibinfo {author} {\bibfnamefont {W.~S.}\ \bibnamefont
  {Lee}}, \bibinfo {author} {\bibfnamefont {J.~J.}\ \bibnamefont {Lee}},
  \bibinfo {author} {\bibfnamefont {E.~A.}\ \bibnamefont {Nowadnick}}, \bibinfo
  {author} {\bibfnamefont {S.}~\bibnamefont {Gerber}}, \bibinfo {author}
  {\bibfnamefont {W.}~\bibnamefont {Tabis}}, \bibinfo {author} {\bibfnamefont
  {S.~W.}\ \bibnamefont {Huang}}, \bibinfo {author} {\bibfnamefont {V.~N.}\
  \bibnamefont {Strocov}}, \bibinfo {author} {\bibfnamefont {E.~M.}\
  \bibnamefont {Motoyama}}, \bibinfo {author} {\bibfnamefont {G.}~\bibnamefont
  {Yu}}, \bibinfo {author} {\bibfnamefont {B.}~\bibnamefont {Moritz}}, \bibinfo
  {author} {\bibfnamefont {H.~Y.}\ \bibnamefont {Huang}}, \bibinfo {author}
  {\bibfnamefont {R.~P.}\ \bibnamefont {Wang}}, \bibinfo {author}
  {\bibfnamefont {Y.~B.}\ \bibnamefont {Huang}}, \bibinfo {author}
  {\bibfnamefont {W.~B.}\ \bibnamefont {Wu}}, \bibinfo {author} {\bibfnamefont
  {C.~T.}\ \bibnamefont {Chen}}, \bibinfo {author} {\bibfnamefont {D.~J.}\
  \bibnamefont {Huang}}, \bibinfo {author} {\bibfnamefont {M.}~\bibnamefont
  {Greven}}, \bibinfo {author} {\bibfnamefont {T.}~\bibnamefont {Schmitt}},
  \bibinfo {author} {\bibfnamefont {Z.~X.}\ \bibnamefont {Shen}}, \ and\
  \bibinfo {author} {\bibfnamefont {T.}~\bibnamefont {Devereaux}},\ }\href
  {\doibase 10.1038/nphys3117} {\bibfield  {journal} {\bibinfo  {journal}
  {Nature Physics}\ }\textbf {\bibinfo {volume} {10}},\ \bibinfo {pages} {883}
  (\bibinfo {year} {2014})}\BibitemShut {NoStop}%
\bibitem [{\citenamefont {da~Silva~Neto}\ \emph {et~al.}(2015)\citenamefont
  {da~Silva~Neto}, \citenamefont {Comin}, \citenamefont {He}, \citenamefont
  {Sutarto}, \citenamefont {Jiang}, \citenamefont {Greene}, \citenamefont
  {Sawatzky},\ and\ \citenamefont {Damascelli}}]{EHdSN_ScienceCDW}%
  \BibitemOpen
  \bibfield  {author} {\bibinfo {author} {\bibfnamefont {E.~H.}\ \bibnamefont
  {da~Silva~Neto}}, \bibinfo {author} {\bibfnamefont {R.}~\bibnamefont
  {Comin}}, \bibinfo {author} {\bibfnamefont {F.}~\bibnamefont {He}}, \bibinfo
  {author} {\bibfnamefont {R.}~\bibnamefont {Sutarto}}, \bibinfo {author}
  {\bibfnamefont {Y.}~\bibnamefont {Jiang}}, \bibinfo {author} {\bibfnamefont
  {R.~L.}\ \bibnamefont {Greene}}, \bibinfo {author} {\bibfnamefont {G.~A.}\
  \bibnamefont {Sawatzky}}, \ and\ \bibinfo {author} {\bibfnamefont
  {A.}~\bibnamefont {Damascelli}},\ }\href {\doibase 10.1126/science.1256441}
  {\bibfield  {journal} {\bibinfo  {journal} {Science}\ }\textbf {\bibinfo
  {volume} {347}},\ \bibinfo {pages} {282} (\bibinfo {year}
  {2015})}\BibitemShut {NoStop}%
\bibitem [{\citenamefont {da~Silva~Neto}\ \emph {et~al.}(2016)\citenamefont
  {da~Silva~Neto}, \citenamefont {Yu}, \citenamefont {Minola}, \citenamefont
  {Sutarto}, \citenamefont {Schierle}, \citenamefont {Boschini}, \citenamefont
  {Zonno}, \citenamefont {Bluschke}, \citenamefont {Higgins}, \citenamefont
  {Li}, \citenamefont {Yu}, \citenamefont {Weschke}, \citenamefont {He},
  \citenamefont {Le~Tacon}, \citenamefont {Greene}, \citenamefont {Greven},
  \citenamefont {Sawatzky}, \citenamefont {Keimer},\ and\ \citenamefont
  {Damascelli}}]{EHdSN_ScienceAdvancesCDW}%
  \BibitemOpen
  \bibfield  {author} {\bibinfo {author} {\bibfnamefont {E.~H.}\ \bibnamefont
  {da~Silva~Neto}}, \bibinfo {author} {\bibfnamefont {B.}~\bibnamefont {Yu}},
  \bibinfo {author} {\bibfnamefont {M.}~\bibnamefont {Minola}}, \bibinfo
  {author} {\bibfnamefont {R.}~\bibnamefont {Sutarto}}, \bibinfo {author}
  {\bibfnamefont {E.}~\bibnamefont {Schierle}}, \bibinfo {author}
  {\bibfnamefont {F.}~\bibnamefont {Boschini}}, \bibinfo {author}
  {\bibfnamefont {M.}~\bibnamefont {Zonno}}, \bibinfo {author} {\bibfnamefont
  {M.}~\bibnamefont {Bluschke}}, \bibinfo {author} {\bibfnamefont
  {J.}~\bibnamefont {Higgins}}, \bibinfo {author} {\bibfnamefont
  {Y.}~\bibnamefont {Li}}, \bibinfo {author} {\bibfnamefont {G.}~\bibnamefont
  {Yu}}, \bibinfo {author} {\bibfnamefont {E.}~\bibnamefont {Weschke}},
  \bibinfo {author} {\bibfnamefont {F.}~\bibnamefont {He}}, \bibinfo {author}
  {\bibfnamefont {M.}~\bibnamefont {Le~Tacon}}, \bibinfo {author}
  {\bibfnamefont {R.~L.}\ \bibnamefont {Greene}}, \bibinfo {author}
  {\bibfnamefont {M.}~\bibnamefont {Greven}}, \bibinfo {author} {\bibfnamefont
  {G.~A.}\ \bibnamefont {Sawatzky}}, \bibinfo {author} {\bibfnamefont
  {B.}~\bibnamefont {Keimer}}, \ and\ \bibinfo {author} {\bibfnamefont
  {A.}~\bibnamefont {Damascelli}},\ }\href {\doibase 10.1126/sciadv.1600782}
  {\bibfield  {journal} {\bibinfo  {journal} {Science Advances}\ }\textbf
  {\bibinfo {volume} {2}} (\bibinfo {year} {2016}),\
  10.1126/sciadv.1600782}\BibitemShut {NoStop}%
\bibitem [{\citenamefont {da~Silva~Neto}\ \emph {et~al.}(2018)\citenamefont
  {da~Silva~Neto}, \citenamefont {Minola}, \citenamefont {Yu}, \citenamefont
  {Tabis}, \citenamefont {Bluschke}, \citenamefont {Unruh}, \citenamefont
  {Suzuki}, \citenamefont {Li}, \citenamefont {Yu}, \citenamefont {Betto},
  \citenamefont {Kummer}, \citenamefont {Yakhou}, \citenamefont {Brookes},
  \citenamefont {Le~Tacon}, \citenamefont {Greven}, \citenamefont {Keimer},\
  and\ \citenamefont {Damascelli}}]{daSilvaNeto_NCCO_2018}%
  \BibitemOpen
  \bibfield  {author} {\bibinfo {author} {\bibfnamefont {E.~H.}\ \bibnamefont
  {da~Silva~Neto}}, \bibinfo {author} {\bibfnamefont {M.}~\bibnamefont
  {Minola}}, \bibinfo {author} {\bibfnamefont {B.}~\bibnamefont {Yu}}, \bibinfo
  {author} {\bibfnamefont {W.}~\bibnamefont {Tabis}}, \bibinfo {author}
  {\bibfnamefont {M.}~\bibnamefont {Bluschke}}, \bibinfo {author}
  {\bibfnamefont {D.}~\bibnamefont {Unruh}}, \bibinfo {author} {\bibfnamefont
  {H.}~\bibnamefont {Suzuki}}, \bibinfo {author} {\bibfnamefont
  {Y.}~\bibnamefont {Li}}, \bibinfo {author} {\bibfnamefont {G.}~\bibnamefont
  {Yu}}, \bibinfo {author} {\bibfnamefont {D.}~\bibnamefont {Betto}}, \bibinfo
  {author} {\bibfnamefont {K.}~\bibnamefont {Kummer}}, \bibinfo {author}
  {\bibfnamefont {F.}~\bibnamefont {Yakhou}}, \bibinfo {author} {\bibfnamefont
  {N.~B.}\ \bibnamefont {Brookes}}, \bibinfo {author} {\bibfnamefont
  {M.}~\bibnamefont {Le~Tacon}}, \bibinfo {author} {\bibfnamefont
  {M.}~\bibnamefont {Greven}}, \bibinfo {author} {\bibfnamefont
  {B.}~\bibnamefont {Keimer}}, \ and\ \bibinfo {author} {\bibfnamefont
  {A.}~\bibnamefont {Damascelli}},\ }\href {\doibase
  10.1103/PhysRevB.98.161114} {\bibfield  {journal} {\bibinfo  {journal} {Phys.
  Rev. B}\ }\textbf {\bibinfo {volume} {98}},\ \bibinfo {pages} {161114}
  (\bibinfo {year} {2018})}\BibitemShut {NoStop}%
\bibitem [{\citenamefont {Hepting}\ \emph {et~al.}(2018)\citenamefont
  {Hepting}, \citenamefont {Chaix}, \citenamefont {Huang}, \citenamefont
  {Fumagalli}, \citenamefont {Peng}, \citenamefont {Moritz}, \citenamefont
  {Kummer}, \citenamefont {Brookes}, \citenamefont {Lee}, \citenamefont
  {Hashimoto}, \citenamefont {Sarkar}, \citenamefont {He}, \citenamefont
  {Rotundu}, \citenamefont {Lee}, \citenamefont {Greene}, \citenamefont
  {Braicovich}, \citenamefont {Ghiringhelli}, \citenamefont {Shen},
  \citenamefont {Devereaux},\ and\ \citenamefont {Lee}}]{Plasmon3D_LCCO}%
  \BibitemOpen
  \bibfield  {author} {\bibinfo {author} {\bibfnamefont {M.}~\bibnamefont
  {Hepting}}, \bibinfo {author} {\bibfnamefont {L.}~\bibnamefont {Chaix}},
  \bibinfo {author} {\bibfnamefont {E.~W.}\ \bibnamefont {Huang}}, \bibinfo
  {author} {\bibfnamefont {R.}~\bibnamefont {Fumagalli}}, \bibinfo {author}
  {\bibfnamefont {Y.~Y.}\ \bibnamefont {Peng}}, \bibinfo {author}
  {\bibfnamefont {B.}~\bibnamefont {Moritz}}, \bibinfo {author} {\bibfnamefont
  {K.}~\bibnamefont {Kummer}}, \bibinfo {author} {\bibfnamefont {N.~B.}\
  \bibnamefont {Brookes}}, \bibinfo {author} {\bibfnamefont {W.~C.}\
  \bibnamefont {Lee}}, \bibinfo {author} {\bibfnamefont {M.}~\bibnamefont
  {Hashimoto}}, \bibinfo {author} {\bibfnamefont {T.}~\bibnamefont {Sarkar}},
  \bibinfo {author} {\bibfnamefont {J.-F.}\ \bibnamefont {He}}, \bibinfo
  {author} {\bibfnamefont {C.~R.}\ \bibnamefont {Rotundu}}, \bibinfo {author}
  {\bibfnamefont {Y.~S.}\ \bibnamefont {Lee}}, \bibinfo {author} {\bibfnamefont
  {R.~L.}\ \bibnamefont {Greene}}, \bibinfo {author} {\bibfnamefont
  {L.}~\bibnamefont {Braicovich}}, \bibinfo {author} {\bibfnamefont
  {G.}~\bibnamefont {Ghiringhelli}}, \bibinfo {author} {\bibfnamefont {Z.-X.}\
  \bibnamefont {Shen}}, \bibinfo {author} {\bibfnamefont {T.~P.}\ \bibnamefont
  {Devereaux}}, \ and\ \bibinfo {author} {\bibfnamefont {W.~S.}\ \bibnamefont
  {Lee}},\ }\href {\doibase 10.1038/s41586-018-0648-3} {\bibfield  {journal}
  {\bibinfo  {journal} {Nature}\ }\textbf {\bibinfo {volume} {563}},\ \bibinfo
  {pages} {374} (\bibinfo {year} {2018})}\BibitemShut {NoStop}%
\bibitem [{\citenamefont {Vilk}\ and\ \citenamefont
  {Tremblay}(1997)}]{TremblayTheorySF}%
  \BibitemOpen
  \bibfield  {author} {\bibinfo {author} {\bibfnamefont {Y.}~\bibnamefont
  {Vilk}}\ and\ \bibinfo {author} {\bibfnamefont {A.-M.}\ \bibnamefont
  {Tremblay}},\ }\href {\doibase 10.1051/jp1:1997135} {\bibfield  {journal}
  {\bibinfo  {journal} {J. Phys. I France}\ }\textbf {\bibinfo {volume} {7}},\
  \bibinfo {pages} {1309} (\bibinfo {year} {1997})}\BibitemShut {NoStop}%
\bibitem [{\citenamefont {Perfetti}\ \emph {et~al.}(2007)\citenamefont
  {Perfetti}, \citenamefont {Loukakos}, \citenamefont {Lisowski}, \citenamefont
  {Bovensiepen}, \citenamefont {Eisaki},\ and\ \citenamefont
  {Wolf}}]{PerfettiPRL2007}%
  \BibitemOpen
  \bibfield  {author} {\bibinfo {author} {\bibfnamefont {L.}~\bibnamefont
  {Perfetti}}, \bibinfo {author} {\bibfnamefont {P.~A.}\ \bibnamefont
  {Loukakos}}, \bibinfo {author} {\bibfnamefont {M.}~\bibnamefont {Lisowski}},
  \bibinfo {author} {\bibfnamefont {U.}~\bibnamefont {Bovensiepen}}, \bibinfo
  {author} {\bibfnamefont {H.}~\bibnamefont {Eisaki}}, \ and\ \bibinfo {author}
  {\bibfnamefont {M.}~\bibnamefont {Wolf}},\ }\href {\doibase
  10.1103/PhysRevLett.99.197001} {\bibfield  {journal} {\bibinfo  {journal}
  {Phys. Rev. Lett.}\ }\textbf {\bibinfo {volume} {99}},\ \bibinfo {pages}
  {197001} (\bibinfo {year} {2007})}\BibitemShut {NoStop}%
\bibitem [{\citenamefont {Dal~Conte}\ \emph {et~al.}(2012)\citenamefont
  {Dal~Conte}, \citenamefont {Giannetti}, \citenamefont {Coslovich},
  \citenamefont {Cilento}, \citenamefont {Bossini}, \citenamefont {Abebaw},
  \citenamefont {Banfi}, \citenamefont {Ferrini}, \citenamefont {Eisaki},
  \citenamefont {Greven}, \citenamefont {Damascelli}, \citenamefont {van~der
  Marel},\ and\ \citenamefont {Parmigiani}}]{ScienceDalConte}%
  \BibitemOpen
  \bibfield  {author} {\bibinfo {author} {\bibfnamefont {S.}~\bibnamefont
  {Dal~Conte}}, \bibinfo {author} {\bibfnamefont {C.}~\bibnamefont
  {Giannetti}}, \bibinfo {author} {\bibfnamefont {G.}~\bibnamefont
  {Coslovich}}, \bibinfo {author} {\bibfnamefont {F.}~\bibnamefont {Cilento}},
  \bibinfo {author} {\bibfnamefont {D.}~\bibnamefont {Bossini}}, \bibinfo
  {author} {\bibfnamefont {T.}~\bibnamefont {Abebaw}}, \bibinfo {author}
  {\bibfnamefont {F.}~\bibnamefont {Banfi}}, \bibinfo {author} {\bibfnamefont
  {G.}~\bibnamefont {Ferrini}}, \bibinfo {author} {\bibfnamefont
  {H.}~\bibnamefont {Eisaki}}, \bibinfo {author} {\bibfnamefont
  {M.}~\bibnamefont {Greven}}, \bibinfo {author} {\bibfnamefont
  {A.}~\bibnamefont {Damascelli}}, \bibinfo {author} {\bibfnamefont
  {D.}~\bibnamefont {van~der Marel}}, \ and\ \bibinfo {author} {\bibfnamefont
  {F.}~\bibnamefont {Parmigiani}},\ }\href {\doibase 10.1126/science.1216765}
  {\bibfield  {journal} {\bibinfo  {journal} {Science}\ }\textbf {\bibinfo
  {volume} {335}},\ \bibinfo {pages} {1600} (\bibinfo {year}
  {2012})}\BibitemShut {NoStop}%
\bibitem [{\citenamefont {Dal~Conte}\ \emph {et~al.}(2015)\citenamefont
  {Dal~Conte}, \citenamefont {Vidmar}, \citenamefont {Golez}, \citenamefont
  {Mierzejewski}, \citenamefont {Soavi}, \citenamefont {Peli}, \citenamefont
  {Banfi}, \citenamefont {Ferrini}, \citenamefont {Comin}, \citenamefont
  {Ludbrook}, \citenamefont {Chauviere}, \citenamefont {Zhigadlo},
  \citenamefont {Eisaki}, \citenamefont {Greven}, \citenamefont {Lupi},
  \citenamefont {Damascelli}, \citenamefont {Brida}, \citenamefont {Capone},
  \citenamefont {Bonca}, \citenamefont {Cerullo},\ and\ \citenamefont
  {Giannetti}}]{NatPhysDalConte}%
  \BibitemOpen
  \bibfield  {author} {\bibinfo {author} {\bibfnamefont {S.}~\bibnamefont
  {Dal~Conte}}, \bibinfo {author} {\bibfnamefont {L.}~\bibnamefont {Vidmar}},
  \bibinfo {author} {\bibfnamefont {D.}~\bibnamefont {Golez}}, \bibinfo
  {author} {\bibfnamefont {M.}~\bibnamefont {Mierzejewski}}, \bibinfo {author}
  {\bibfnamefont {G.}~\bibnamefont {Soavi}}, \bibinfo {author} {\bibfnamefont
  {S.}~\bibnamefont {Peli}}, \bibinfo {author} {\bibfnamefont {F.}~\bibnamefont
  {Banfi}}, \bibinfo {author} {\bibfnamefont {G.}~\bibnamefont {Ferrini}},
  \bibinfo {author} {\bibfnamefont {R.}~\bibnamefont {Comin}}, \bibinfo
  {author} {\bibfnamefont {B.~M.}\ \bibnamefont {Ludbrook}}, \bibinfo {author}
  {\bibfnamefont {L.}~\bibnamefont {Chauviere}}, \bibinfo {author}
  {\bibfnamefont {N.~D.}\ \bibnamefont {Zhigadlo}}, \bibinfo {author}
  {\bibfnamefont {H.}~\bibnamefont {Eisaki}}, \bibinfo {author} {\bibfnamefont
  {M.}~\bibnamefont {Greven}}, \bibinfo {author} {\bibfnamefont
  {S.}~\bibnamefont {Lupi}}, \bibinfo {author} {\bibfnamefont {A.}~\bibnamefont
  {Damascelli}}, \bibinfo {author} {\bibfnamefont {D.}~\bibnamefont {Brida}},
  \bibinfo {author} {\bibfnamefont {M.}~\bibnamefont {Capone}}, \bibinfo
  {author} {\bibfnamefont {J.}~\bibnamefont {Bonca}}, \bibinfo {author}
  {\bibfnamefont {G.}~\bibnamefont {Cerullo}}, \ and\ \bibinfo {author}
  {\bibfnamefont {C.}~\bibnamefont {Giannetti}},\ }\href {\doibase
  10.1038/nphys3265} {\bibfield  {journal} {\bibinfo  {journal} {Nature
  Physics}\ }\textbf {\bibinfo {volume} {11}},\ \bibinfo {pages} {421}
  (\bibinfo {year} {2015})}\BibitemShut {NoStop}%
\bibitem [{\citenamefont {Giannetti}\ \emph {et~al.}(2016)\citenamefont
  {Giannetti}, \citenamefont {Capone}, \citenamefont {Fausti}, \citenamefont
  {Fabrizio}, \citenamefont {Parmigiani},\ and\ \citenamefont
  {Mihailovic}}]{ReviewGiannetti}%
  \BibitemOpen
  \bibfield  {author} {\bibinfo {author} {\bibfnamefont {C.}~\bibnamefont
  {Giannetti}}, \bibinfo {author} {\bibfnamefont {M.}~\bibnamefont {Capone}},
  \bibinfo {author} {\bibfnamefont {D.}~\bibnamefont {Fausti}}, \bibinfo
  {author} {\bibfnamefont {M.}~\bibnamefont {Fabrizio}}, \bibinfo {author}
  {\bibfnamefont {F.}~\bibnamefont {Parmigiani}}, \ and\ \bibinfo {author}
  {\bibfnamefont {D.}~\bibnamefont {Mihailovic}},\ }\href {\doibase
  10.1080/00018732.2016.1194044} {\bibfield  {journal} {\bibinfo  {journal}
  {Advances in Physics}\ }\textbf {\bibinfo {volume} {65}},\ \bibinfo {pages}
  {58} (\bibinfo {year} {2016})}\BibitemShut {NoStop}%
\bibitem [{\citenamefont {Hinton}\ \emph {et~al.}(2013)\citenamefont {Hinton},
  \citenamefont {Koralek}, \citenamefont {Yu}, \citenamefont {Motoyama},
  \citenamefont {Lu}, \citenamefont {Vishwanath}, \citenamefont {Greven},\ and\
  \citenamefont {Orenstein}}]{Hinton_TR_NCCO}%
  \BibitemOpen
  \bibfield  {author} {\bibinfo {author} {\bibfnamefont {J.~P.}\ \bibnamefont
  {Hinton}}, \bibinfo {author} {\bibfnamefont {J.~D.}\ \bibnamefont {Koralek}},
  \bibinfo {author} {\bibfnamefont {G.}~\bibnamefont {Yu}}, \bibinfo {author}
  {\bibfnamefont {E.~M.}\ \bibnamefont {Motoyama}}, \bibinfo {author}
  {\bibfnamefont {Y.~M.}\ \bibnamefont {Lu}}, \bibinfo {author} {\bibfnamefont
  {A.}~\bibnamefont {Vishwanath}}, \bibinfo {author} {\bibfnamefont
  {M.}~\bibnamefont {Greven}}, \ and\ \bibinfo {author} {\bibfnamefont
  {J.}~\bibnamefont {Orenstein}},\ }\href {\doibase
  10.1103/PhysRevLett.110.217002} {\bibfield  {journal} {\bibinfo  {journal}
  {Phys. Rev. Lett.}\ }\textbf {\bibinfo {volume} {110}},\ \bibinfo {pages}
  {217002} (\bibinfo {year} {2013})}\BibitemShut {NoStop}%
\bibitem [{\citenamefont {Vishik}\ \emph {et~al.}(2017)\citenamefont {Vishik},
  \citenamefont {Mahmood}, \citenamefont {Alpichshev}, \citenamefont {Gedik},
  \citenamefont {Higgins},\ and\ \citenamefont {Greene}}]{Vishik_TR_LCCO}%
  \BibitemOpen
  \bibfield  {author} {\bibinfo {author} {\bibfnamefont {I.~M.}\ \bibnamefont
  {Vishik}}, \bibinfo {author} {\bibfnamefont {F.}~\bibnamefont {Mahmood}},
  \bibinfo {author} {\bibfnamefont {Z.}~\bibnamefont {Alpichshev}}, \bibinfo
  {author} {\bibfnamefont {N.}~\bibnamefont {Gedik}}, \bibinfo {author}
  {\bibfnamefont {J.}~\bibnamefont {Higgins}}, \ and\ \bibinfo {author}
  {\bibfnamefont {R.~L.}\ \bibnamefont {Greene}},\ }\href {\doibase
  10.1103/PhysRevB.95.115125} {\bibfield  {journal} {\bibinfo  {journal} {Phys.
  Rev. B}\ }\textbf {\bibinfo {volume} {95}},\ \bibinfo {pages} {115125}
  (\bibinfo {year} {2017})}\BibitemShut {NoStop}%
\bibitem [{\citenamefont {Graf}\ \emph {et~al.}(2011)\citenamefont {Graf},
  \citenamefont {Jozwiak}, \citenamefont {Smallwood}, \citenamefont {Eisaki},
  \citenamefont {Kaindl}, \citenamefont {Lee},\ and\ \citenamefont
  {Lanzara}}]{NodePeakLanzara}%
  \BibitemOpen
  \bibfield  {author} {\bibinfo {author} {\bibfnamefont {J.}~\bibnamefont
  {Graf}}, \bibinfo {author} {\bibfnamefont {C.}~\bibnamefont {Jozwiak}},
  \bibinfo {author} {\bibfnamefont {C.~L.}\ \bibnamefont {Smallwood}}, \bibinfo
  {author} {\bibfnamefont {H.}~\bibnamefont {Eisaki}}, \bibinfo {author}
  {\bibfnamefont {R.~A.}\ \bibnamefont {Kaindl}}, \bibinfo {author}
  {\bibfnamefont {D.-H.}\ \bibnamefont {Lee}}, \ and\ \bibinfo {author}
  {\bibfnamefont {A.}~\bibnamefont {Lanzara}},\ }\href {\doibase
  10.1038/nphys2027} {\bibfield  {journal} {\bibinfo  {journal} {Nature
  Physics}\ }\textbf {\bibinfo {volume} {7}},\ \bibinfo {pages} {805} (\bibinfo
  {year} {2011})}\BibitemShut {NoStop}%
\bibitem [{\citenamefont {Markiewicz}\ \emph {et~al.}(2005)\citenamefont
  {Markiewicz}, \citenamefont {Sahrakorpi}, \citenamefont {Lindroos},
  \citenamefont {Lin},\ and\ \citenamefont {Bansil}}]{TB_NCCO}%
  \BibitemOpen
  \bibfield  {author} {\bibinfo {author} {\bibfnamefont {R.~S.}\ \bibnamefont
  {Markiewicz}}, \bibinfo {author} {\bibfnamefont {S.}~\bibnamefont
  {Sahrakorpi}}, \bibinfo {author} {\bibfnamefont {M.}~\bibnamefont
  {Lindroos}}, \bibinfo {author} {\bibfnamefont {H.}~\bibnamefont {Lin}}, \
  and\ \bibinfo {author} {\bibfnamefont {A.}~\bibnamefont {Bansil}},\ }\href
  {\doibase 10.1103/PhysRevB.72.054519} {\bibfield  {journal} {\bibinfo
  {journal} {Phys. Rev. B}\ }\textbf {\bibinfo {volume} {72}},\ \bibinfo
  {pages} {054519} (\bibinfo {year} {2005})}\BibitemShut {NoStop}%
\bibitem [{\citenamefont {Damascelli}\ \emph {et~al.}(2003)\citenamefont
  {Damascelli}, \citenamefont {Hussain},\ and\ \citenamefont
  {Shen}}]{ReviewDamascelli}%
  \BibitemOpen
  \bibfield  {author} {\bibinfo {author} {\bibfnamefont {A.}~\bibnamefont
  {Damascelli}}, \bibinfo {author} {\bibfnamefont {Z.}~\bibnamefont {Hussain}},
  \ and\ \bibinfo {author} {\bibfnamefont {Z.-X.}\ \bibnamefont {Shen}},\
  }\href {\doibase 10.1103/RevModPhys.75.473} {\bibfield  {journal} {\bibinfo
  {journal} {Rev. Mod. Phys.}\ }\textbf {\bibinfo {volume} {75}},\ \bibinfo
  {pages} {473} (\bibinfo {year} {2003})}\BibitemShut {NoStop}%
\bibitem [{\citenamefont {He}\ \emph {et~al.}(2019)\citenamefont {He},
  \citenamefont {Rotundu}, \citenamefont {Scheurer}, \citenamefont {He},
  \citenamefont {Hashimoto}, \citenamefont {Xu}, \citenamefont {Wang},
  \citenamefont {Huang}, \citenamefont {Jia}, \citenamefont {Chen},
  \citenamefont {Moritz}, \citenamefont {Lu}, \citenamefont {Lee},
  \citenamefont {Devereaux},\ and\ \citenamefont {Shen}}]{Shen_NCCO_arXiv}%
  \BibitemOpen
  \bibfield  {author} {\bibinfo {author} {\bibfnamefont {J.}~\bibnamefont
  {He}}, \bibinfo {author} {\bibfnamefont {C.~R.}\ \bibnamefont {Rotundu}},
  \bibinfo {author} {\bibfnamefont {M.~S.}\ \bibnamefont {Scheurer}}, \bibinfo
  {author} {\bibfnamefont {Y.}~\bibnamefont {He}}, \bibinfo {author}
  {\bibfnamefont {M.}~\bibnamefont {Hashimoto}}, \bibinfo {author}
  {\bibfnamefont {K.-J.}\ \bibnamefont {Xu}}, \bibinfo {author} {\bibfnamefont
  {Y.}~\bibnamefont {Wang}}, \bibinfo {author} {\bibfnamefont {E.~W.}\
  \bibnamefont {Huang}}, \bibinfo {author} {\bibfnamefont {T.}~\bibnamefont
  {Jia}}, \bibinfo {author} {\bibfnamefont {S.}~\bibnamefont {Chen}}, \bibinfo
  {author} {\bibfnamefont {B.}~\bibnamefont {Moritz}}, \bibinfo {author}
  {\bibfnamefont {D.}~\bibnamefont {Lu}}, \bibinfo {author} {\bibfnamefont
  {Y.~S.}\ \bibnamefont {Lee}}, \bibinfo {author} {\bibfnamefont {T.~P.}\
  \bibnamefont {Devereaux}}, \ and\ \bibinfo {author} {\bibfnamefont {Z.-X.}\
  \bibnamefont {Shen}},\ }\href {\doibase 10.1073/pnas.1816121116} {\bibfield
  {journal} {\bibinfo  {journal} {Proceedings of the National Academy of
  Sciences}\ }\textbf {\bibinfo {volume} {116}},\ \bibinfo {pages} {3449}
  (\bibinfo {year} {2019})}\BibitemShut {NoStop}%
\bibitem [{\citenamefont {Norman}\ \emph {et~al.}(1998)\citenamefont {Norman},
  \citenamefont {Ding}, \citenamefont {Randeria}, \citenamefont {Campuzano},
  \citenamefont {Yokoya}, \citenamefont {Takeuchi}, \citenamefont {Takahashi},
  \citenamefont {Mochiku}, \citenamefont {Kadowaki}, \citenamefont
  {Guptasarma},\ and\ \citenamefont {Hinks}}]{NormanSymmetrizedEDCNature}%
  \BibitemOpen
  \bibfield  {author} {\bibinfo {author} {\bibfnamefont {M.~R.}\ \bibnamefont
  {Norman}}, \bibinfo {author} {\bibfnamefont {H.}~\bibnamefont {Ding}},
  \bibinfo {author} {\bibfnamefont {M.}~\bibnamefont {Randeria}}, \bibinfo
  {author} {\bibfnamefont {J.~C.}\ \bibnamefont {Campuzano}}, \bibinfo {author}
  {\bibfnamefont {T.}~\bibnamefont {Yokoya}}, \bibinfo {author} {\bibfnamefont
  {T.}~\bibnamefont {Takeuchi}}, \bibinfo {author} {\bibfnamefont
  {T.}~\bibnamefont {Takahashi}}, \bibinfo {author} {\bibfnamefont
  {T.}~\bibnamefont {Mochiku}}, \bibinfo {author} {\bibfnamefont
  {K.}~\bibnamefont {Kadowaki}}, \bibinfo {author} {\bibfnamefont
  {P.}~\bibnamefont {Guptasarma}}, \ and\ \bibinfo {author} {\bibfnamefont
  {D.~G.}\ \bibnamefont {Hinks}},\ }\href {\doibase 10.1038/32366} {\bibfield
  {journal} {\bibinfo  {journal} {Nature}\ }\textbf {\bibinfo {volume} {392}},\
  \bibinfo {pages} {157} (\bibinfo {year} {1998})}\BibitemShut {NoStop}%
\bibitem [{\citenamefont {Miller}\ \emph {et~al.}(2018)\citenamefont {Miller},
  \citenamefont {Zhang}, \citenamefont {Ma}, \citenamefont {Eisaki},
  \citenamefont {Moore},\ and\ \citenamefont {Lanzara}}]{MillerKink2212}%
  \BibitemOpen
  \bibfield  {author} {\bibinfo {author} {\bibfnamefont {T.~L.}\ \bibnamefont
  {Miller}}, \bibinfo {author} {\bibfnamefont {W.}~\bibnamefont {Zhang}},
  \bibinfo {author} {\bibfnamefont {J.}~\bibnamefont {Ma}}, \bibinfo {author}
  {\bibfnamefont {H.}~\bibnamefont {Eisaki}}, \bibinfo {author} {\bibfnamefont
  {J.~E.}\ \bibnamefont {Moore}}, \ and\ \bibinfo {author} {\bibfnamefont
  {A.}~\bibnamefont {Lanzara}},\ }\href {\doibase 10.1103/PhysRevB.97.134517}
  {\bibfield  {journal} {\bibinfo  {journal} {Phys. Rev. B}\ }\textbf {\bibinfo
  {volume} {97}},\ \bibinfo {pages} {134517} (\bibinfo {year}
  {2018})}\BibitemShut {NoStop}%
\bibitem [{\citenamefont {Boschini}\ \emph {et~al.}(2018)\citenamefont
  {Boschini}, \citenamefont {da~Silva~Neto}, \citenamefont {Razzoli},
  \citenamefont {Zonno}, \citenamefont {Peli}, \citenamefont {Day},
  \citenamefont {Michiardi}, \citenamefont {Schneider}, \citenamefont
  {Zwartsenberg}, \citenamefont {Nigge}, \citenamefont {Zhong}, \citenamefont
  {Schneeloch}, \citenamefont {Gu}, \citenamefont {Zhdanovich}, \citenamefont
  {Mills}, \citenamefont {Levy}, \citenamefont {Jones},\ and\ \citenamefont
  {Damascelli}}]{Boschini2018}%
  \BibitemOpen
  \bibfield  {author} {\bibinfo {author} {\bibfnamefont {F.}~\bibnamefont
  {Boschini}}, \bibinfo {author} {\bibfnamefont {E.~H.}\ \bibnamefont
  {da~Silva~Neto}}, \bibinfo {author} {\bibfnamefont {E.}~\bibnamefont
  {Razzoli}}, \bibinfo {author} {\bibfnamefont {M.}~\bibnamefont {Zonno}},
  \bibinfo {author} {\bibfnamefont {S.}~\bibnamefont {Peli}}, \bibinfo {author}
  {\bibfnamefont {R.~P.}\ \bibnamefont {Day}}, \bibinfo {author} {\bibfnamefont
  {M.}~\bibnamefont {Michiardi}}, \bibinfo {author} {\bibfnamefont
  {M.}~\bibnamefont {Schneider}}, \bibinfo {author} {\bibfnamefont
  {B.}~\bibnamefont {Zwartsenberg}}, \bibinfo {author} {\bibfnamefont
  {P.}~\bibnamefont {Nigge}}, \bibinfo {author} {\bibfnamefont {R.~D.}\
  \bibnamefont {Zhong}}, \bibinfo {author} {\bibfnamefont {J.}~\bibnamefont
  {Schneeloch}}, \bibinfo {author} {\bibfnamefont {G.~D.}\ \bibnamefont {Gu}},
  \bibinfo {author} {\bibfnamefont {S.}~\bibnamefont {Zhdanovich}}, \bibinfo
  {author} {\bibfnamefont {A.~K.}\ \bibnamefont {Mills}}, \bibinfo {author}
  {\bibfnamefont {G.}~\bibnamefont {Levy}}, \bibinfo {author} {\bibfnamefont
  {C.}~\bibnamefont {Jones}, \bibfnamefont {D.~J.~andGiannetti}}, \ and\
  \bibinfo {author} {\bibfnamefont {A.}~\bibnamefont {Damascelli}},\ }\href
  {\doibase 10.1038/s41563-018-0045-1} {\bibfield  {journal} {\bibinfo
  {journal} {Nature Materials}\ }\textbf {\bibinfo {volume} {17}},\ \bibinfo
  {pages} {416} (\bibinfo {year} {2018})}\BibitemShut {NoStop}%
\bibitem [{\citenamefont {Parham}\ \emph {et~al.}(2017)\citenamefont {Parham},
  \citenamefont {Li}, \citenamefont {Nummy}, \citenamefont {Waugh},
  \citenamefont {Zhou}, \citenamefont {Griffith}, \citenamefont {Schneeloch},
  \citenamefont {Zhong}, \citenamefont {Gu},\ and\ \citenamefont
  {Dessau}}]{DessauPRX}%
  \BibitemOpen
  \bibfield  {author} {\bibinfo {author} {\bibfnamefont {S.}~\bibnamefont
  {Parham}}, \bibinfo {author} {\bibfnamefont {H.}~\bibnamefont {Li}}, \bibinfo
  {author} {\bibfnamefont {T.~J.}\ \bibnamefont {Nummy}}, \bibinfo {author}
  {\bibfnamefont {J.~A.}\ \bibnamefont {Waugh}}, \bibinfo {author}
  {\bibfnamefont {X.~Q.}\ \bibnamefont {Zhou}}, \bibinfo {author}
  {\bibfnamefont {J.}~\bibnamefont {Griffith}}, \bibinfo {author}
  {\bibfnamefont {J.}~\bibnamefont {Schneeloch}}, \bibinfo {author}
  {\bibfnamefont {R.~D.}\ \bibnamefont {Zhong}}, \bibinfo {author}
  {\bibfnamefont {G.~D.}\ \bibnamefont {Gu}}, \ and\ \bibinfo {author}
  {\bibfnamefont {D.~S.}\ \bibnamefont {Dessau}},\ }\href {\doibase
  10.1103/PhysRevX.7.041013} {\bibfield  {journal} {\bibinfo  {journal} {Phys.
  Rev. X}\ }\textbf {\bibinfo {volume} {7}},\ \bibinfo {pages} {041013}
  (\bibinfo {year} {2017})}\BibitemShut {NoStop}%
\bibitem [{\citenamefont {Kyung}\ \emph {et~al.}(2004)\citenamefont {Kyung},
  \citenamefont {Hankevych}, \citenamefont {Dar\'e},\ and\ \citenamefont
  {Tremblay}}]{KyungPRL2004}%
  \BibitemOpen
  \bibfield  {author} {\bibinfo {author} {\bibfnamefont {B.}~\bibnamefont
  {Kyung}}, \bibinfo {author} {\bibfnamefont {V.}~\bibnamefont {Hankevych}},
  \bibinfo {author} {\bibfnamefont {A.-M.}\ \bibnamefont {Dar\'e}}, \ and\
  \bibinfo {author} {\bibfnamefont {A.-M.~S.}\ \bibnamefont {Tremblay}},\
  }\href {\doibase 10.1103/PhysRevLett.93.147004} {\bibfield  {journal}
  {\bibinfo  {journal} {Phys. Rev. Lett.}\ }\textbf {\bibinfo {volume} {93}},\
  \bibinfo {pages} {147004} (\bibinfo {year} {2004})}\BibitemShut {NoStop}%
\bibitem [{\citenamefont {Li}\ and\ \citenamefont
  {Yao}(2018)}]{TheoryPCCO_fillingPG}%
  \BibitemOpen
  \bibfield  {author} {\bibinfo {author} {\bibfnamefont {T.}~\bibnamefont
  {Li}}\ and\ \bibinfo {author} {\bibfnamefont {D.-W.}\ \bibnamefont {Yao}},\
  }\href {\doibase 10.1209/0295-5075/124/47001} {\bibfield  {journal} {\bibinfo
   {journal} {{EPL} (Europhysics Letters)}\ }\textbf {\bibinfo {volume}
  {124}},\ \bibinfo {pages} {47001} (\bibinfo {year} {2018})}\BibitemShut
  {NoStop}%
\bibitem [{\citenamefont {Helm}\ \emph {et~al.}(2009)\citenamefont {Helm},
  \citenamefont {Kartsovnik}, \citenamefont {Bartkowiak}, \citenamefont
  {Bittner}, \citenamefont {Lambacher}, \citenamefont {Erb}, \citenamefont
  {Wosnitza},\ and\ \citenamefont {Gross}}]{QuantumOscillationsNCCO}%
  \BibitemOpen
  \bibfield  {author} {\bibinfo {author} {\bibfnamefont {T.}~\bibnamefont
  {Helm}}, \bibinfo {author} {\bibfnamefont {M.~V.}\ \bibnamefont
  {Kartsovnik}}, \bibinfo {author} {\bibfnamefont {M.}~\bibnamefont
  {Bartkowiak}}, \bibinfo {author} {\bibfnamefont {N.}~\bibnamefont {Bittner}},
  \bibinfo {author} {\bibfnamefont {M.}~\bibnamefont {Lambacher}}, \bibinfo
  {author} {\bibfnamefont {A.}~\bibnamefont {Erb}}, \bibinfo {author}
  {\bibfnamefont {J.}~\bibnamefont {Wosnitza}}, \ and\ \bibinfo {author}
  {\bibfnamefont {R.}~\bibnamefont {Gross}},\ }\href {\doibase
  10.1103/PhysRevLett.103.157002} {\bibfield  {journal} {\bibinfo  {journal}
  {Phys. Rev. Lett.}\ }\textbf {\bibinfo {volume} {103}},\ \bibinfo {pages}
  {157002} (\bibinfo {year} {2009})}\BibitemShut {NoStop}%
\bibitem [{\citenamefont {Cyr-Choini\`ere}\ \emph {et~al.}(2017)\citenamefont
  {Cyr-Choini\`ere}, \citenamefont {Badoux}, \citenamefont {Grissonnanche},
  \citenamefont {Michon}, \citenamefont {Afshar}, \citenamefont {Fortier},
  \citenamefont {LeBoeuf}, \citenamefont {Graf}, \citenamefont {Day},
  \citenamefont {Bonn}, \citenamefont {Hardy}, \citenamefont {Liang},
  \citenamefont {Doiron-Leyraud},\ and\ \citenamefont
  {Taillefer}}]{TailleferPRX2017}%
  \BibitemOpen
  \bibfield  {author} {\bibinfo {author} {\bibfnamefont {O.}~\bibnamefont
  {Cyr-Choini\`ere}}, \bibinfo {author} {\bibfnamefont {S.}~\bibnamefont
  {Badoux}}, \bibinfo {author} {\bibfnamefont {G.}~\bibnamefont
  {Grissonnanche}}, \bibinfo {author} {\bibfnamefont {B.}~\bibnamefont
  {Michon}}, \bibinfo {author} {\bibfnamefont {S.~A.~A.}\ \bibnamefont
  {Afshar}}, \bibinfo {author} {\bibfnamefont {S.}~\bibnamefont {Fortier}},
  \bibinfo {author} {\bibfnamefont {D.}~\bibnamefont {LeBoeuf}}, \bibinfo
  {author} {\bibfnamefont {D.}~\bibnamefont {Graf}}, \bibinfo {author}
  {\bibfnamefont {J.}~\bibnamefont {Day}}, \bibinfo {author} {\bibfnamefont
  {D.~A.}\ \bibnamefont {Bonn}}, \bibinfo {author} {\bibfnamefont {W.~N.}\
  \bibnamefont {Hardy}}, \bibinfo {author} {\bibfnamefont {R.}~\bibnamefont
  {Liang}}, \bibinfo {author} {\bibfnamefont {N.}~\bibnamefont
  {Doiron-Leyraud}}, \ and\ \bibinfo {author} {\bibfnamefont {L.}~\bibnamefont
  {Taillefer}},\ }\href {\doibase 10.1103/PhysRevX.7.031042} {\bibfield
  {journal} {\bibinfo  {journal} {Phys. Rev. X}\ }\textbf {\bibinfo {volume}
  {7}},\ \bibinfo {pages} {031042} (\bibinfo {year} {2017})}\BibitemShut
  {NoStop}%
\bibitem [{\citenamefont {Lambacher}\ \emph {et~al.}(2010)\citenamefont
  {Lambacher}, \citenamefont {Helm}, \citenamefont {Kartsovnik},\ and\
  \citenamefont {Erb}}]{Erb2010}%
  \BibitemOpen
  \bibfield  {author} {\bibinfo {author} {\bibfnamefont {M.}~\bibnamefont
  {Lambacher}}, \bibinfo {author} {\bibfnamefont {T.}~\bibnamefont {Helm}},
  \bibinfo {author} {\bibfnamefont {M.}~\bibnamefont {Kartsovnik}}, \ and\
  \bibinfo {author} {\bibfnamefont {A.}~\bibnamefont {Erb}},\ }\href {\doibase
  10.1140/epjst/e2010-01297-8} {\bibfield  {journal} {\bibinfo  {journal} {The
  European Physical Journal Special Topics}\ }\textbf {\bibinfo {volume}
  {188}},\ \bibinfo {pages} {61} (\bibinfo {year} {2010})}\BibitemShut
  {NoStop}%
\bibitem [{\citenamefont {Werfel}\ \emph {et~al.}(2015)\citenamefont {Werfel},
  \citenamefont {Erb}, \citenamefont {Fuchs}, \citenamefont {Krabbes},
  \citenamefont {Canders}, \citenamefont {Wördenweber}, \citenamefont {Meyer},
  \citenamefont {Fritzsch}, \citenamefont {Anders}, \citenamefont {Schmelz},
  \citenamefont {Kunert}, \citenamefont {Oelsner}, \citenamefont {Tanabe},
  \citenamefont {Shen}, \citenamefont {Jiang}, \citenamefont {Hellstrom},
  \citenamefont {Paranthaman}, \citenamefont {Aytug}, \citenamefont {Stan},
  \citenamefont {Jia}, \citenamefont {Cantoni}, \citenamefont {Bottura},
  \citenamefont {Luongo}, \citenamefont {Kaiser}, \citenamefont {Schroeder},\
  and\ \citenamefont {Kade}}]{BookErb}%
  \BibitemOpen
  \bibfield  {author} {\bibinfo {author} {\bibfnamefont {F.~N.}\ \bibnamefont
  {Werfel}}, \bibinfo {author} {\bibfnamefont {A.}~\bibnamefont {Erb}},
  \bibinfo {author} {\bibfnamefont {G.}~\bibnamefont {Fuchs}}, \bibinfo
  {author} {\bibfnamefont {G.}~\bibnamefont {Krabbes}}, \bibinfo {author}
  {\bibfnamefont {W.-R.}\ \bibnamefont {Canders}}, \bibinfo {author}
  {\bibfnamefont {R.}~\bibnamefont {Wördenweber}}, \bibinfo {author}
  {\bibfnamefont {H.-G.}\ \bibnamefont {Meyer}}, \bibinfo {author}
  {\bibfnamefont {L.}~\bibnamefont {Fritzsch}}, \bibinfo {author}
  {\bibfnamefont {S.}~\bibnamefont {Anders}}, \bibinfo {author} {\bibfnamefont
  {M.}~\bibnamefont {Schmelz}}, \bibinfo {author} {\bibfnamefont
  {J.}~\bibnamefont {Kunert}}, \bibinfo {author} {\bibfnamefont
  {G.}~\bibnamefont {Oelsner}}, \bibinfo {author} {\bibfnamefont
  {K.}~\bibnamefont {Tanabe}}, \bibinfo {author} {\bibfnamefont
  {T.}~\bibnamefont {Shen}}, \bibinfo {author} {\bibfnamefont {J.}~\bibnamefont
  {Jiang}}, \bibinfo {author} {\bibfnamefont {E.}~\bibnamefont {Hellstrom}},
  \bibinfo {author} {\bibfnamefont {M.~P.}\ \bibnamefont {Paranthaman}},
  \bibinfo {author} {\bibfnamefont {T.}~\bibnamefont {Aytug}}, \bibinfo
  {author} {\bibfnamefont {L.}~\bibnamefont {Stan}}, \bibinfo {author}
  {\bibfnamefont {Q.}~\bibnamefont {Jia}}, \bibinfo {author} {\bibfnamefont
  {C.}~\bibnamefont {Cantoni}}, \bibinfo {author} {\bibfnamefont
  {L.}~\bibnamefont {Bottura}}, \bibinfo {author} {\bibfnamefont
  {C.}~\bibnamefont {Luongo}}, \bibinfo {author} {\bibfnamefont
  {G.}~\bibnamefont {Kaiser}}, \bibinfo {author} {\bibfnamefont
  {G.}~\bibnamefont {Schroeder}}, \ and\ \bibinfo {author} {\bibfnamefont
  {A.}~\bibnamefont {Kade}},\ }\enquote {\bibinfo {title} {Technology,
  preparation, and characterization},}\ in\ \href {\doibase
  10.1002/9783527670635.ch3} {\emph {\bibinfo {booktitle} {Applied
  Superconductivity}}}\ (\bibinfo  {publisher} {Wiley-Blackwell},\ \bibinfo
  {year} {2015})\ Chap.~\bibinfo {chapter} {3}, pp.\ \bibinfo {pages}
  {193--402}\BibitemShut {NoStop}%
\end{thebibliography}
\providecommand{\noopsort}[1]{}\providecommand{\singleletter}[1]{#1}

\newpage 
\linespread{1}

\end{document}